\newif\ifhassvjourthree
\IfFileExists{svjour3.cls}{\hassvjourthreetrue}{\hassvjourthreefalse}
\ifhassvjourthree
  \documentclass[smallextended]{svjour3}
\else
  \documentclass[11pt,a4paper]{article}
  \newcommand{\smartqed}{}
  \newcommand{\keywords}[1]{\par\medskip\noindent\textbf{Keywords} #1}
  \newcommand{\institute}[1]{}\newcommand{\at}{\\}
  \newcommand{\email}[1]{\texttt{#1}}
  \newcommand{\titlerunning}[1]{}\newcommand{\authorrunning}[1]{}
  \usepackage[a4paper,textwidth=16.5cm,textheight=23.5cm,centering]{geometry}
\fi
\smartqed

\usepackage[T1]{fontenc}\usepackage[utf8]{inputenc}
\usepackage{amsmath,amssymb,amsfonts,amsthm,mathtools}
\usepackage{graphicx,booktabs,multirow,xcolor,enumitem}
\providecommand{\dg}{\ensuremath{{}^{\circ}}}
\providecommand{\Ry}{R_y}
\providecommand{\TV}{\operatorname{TV}}
\providecommand{\Fid}{\operatorname{Fid}}

\providecommand{\FWER}{\mathrm{FWER}}
\providecommand{\FDR}{\mathrm{FDR}}
\providecommand{\E}{\mathbb{E}}
\providecommand{\PR}{\mathbb{P}}

\newcommand{\genNsigHolm}{3}
\newcommand{\genNsigBH}{4}
\newcommand{\genSigHolmList}{D0, D5, D6}
\newcommand{\genSigBHList}{D0, D3, D5, D6}
\newcommand{\genNsigBY}{3}
\newcommand{\genSigBYList}{D0, D5, D6}
\newcommand{\genBYconst}{2.5929}
\newcommand{\genNreps}{25}
\newcommand{\genInterEst}{+0.0625}
\newcommand{\genInterLo}{+0.0306}
\newcommand{\genInterHi}{+0.0944}
\newcommand{\genInterP}{4.8\times10^{-4}}
\newcommand{\genCmpCtlAB}{+0.0395}

\newcommand{\genCmpCtlGates}{21}
\newcommand{\genCmpDirAB}{+0.0146}

\newcommand{\genCmpAprGates}{19}
\newcommand{\genCmpRedAB}{-0.0229}

\newcommand{\genDiffCtlEst}{+0.0410}
\newcommand{\genDiffCtlLo}{+0.0138}
\newcommand{\genDiffCtlHi}{+0.0681}
\newcommand{\genDiffCtlZ}{2.95}
\newcommand{\genDiffCtlP}{0.003}
\newcommand{\genDiffRedEst}{-0.0215}
\newcommand{\genDiffRedLo}{-0.0441}
\newcommand{\genDiffRedHi}{+0.0011}
\newcommand{\genDiffRedZ}{-1.86}
\newcommand{\genDiffRedP}{0.063}

\newcommand{\genTwoDayShiftA}{+0.0347}
\newcommand{\genTwoDayPA}{0.014}
\newcommand{\genTwoDayShiftB}{+0.0596}
\newcommand{\genTwoDayPB}{8.9\times10^{-6}}
\newcommand{\genTwoDayFiveOct}{0.8665}
\newcommand{\genTwoDaySevenOct}{0.9136}
\newcommand{\genTwoDayShift}{+0.0471}
\newcommand{\genTwoDayPooledP}{1.3\times10^{-6}}
\newcommand{\genTwoDayDiffEst}{-0.0249}
\newcommand{\genTwoDayDiffLo}{-0.0572}
\newcommand{\genTwoDayDiffHi}{+0.0073}
\newcommand{\genTwoDayDiffZ}{-1.52}
\newcommand{\genTwoDayDiffP}{0.130}
\newcommand{\genCorrLambda}{0.564}
\newcommand{\genCorrRtwo}{0.74}
\newcommand{\genCorrResidMax}{0.357}
\newcommand{\genCorrLamMin}{-0.12}
\newcommand{\genCorrLamMax}{+1.22}
\newcommand{\genCorrZeroN}{20}
\newcommand{\genCorrZeroMax}{0.217}
\newcommand{\genCorrZeroRms}{0.072}
\newcommand{\genCorrDzeroZZ}{+0.217}
\newcommand{\genCorrDzeroF}{0.962}
\newcommand{\genNbalancedTargets}{4}

\newcommand{\genMargDevDthree}{0.029}
\newcommand{\genMargDevDfour}{0.017}

\newcommand{\genCorrRatioDthree}{41}

\newcommand{\genCorrRatioDfour}{22}
\newcommand{\genCorrTgtDfour}{-0.720}
\newcommand{\genCorrObsDfour}{-0.160}
\newcommand{\genLenientFid}{0.908}
\newcommand{\genLenientTV}{0.268}
\newcommand{\genCorrRedAEst}{-0.113}
\newcommand{\genCorrRedALo}{-0.205}
\newcommand{\genCorrRedAHi}{-0.022}
\newcommand{\genCorrRedAP}{0.017}
\newcommand{\genCorrRedARes}{+0.046}
\newcommand{\genCorrRedBEst}{-0.207}
\newcommand{\genCorrRedBLo}{-0.278}
\newcommand{\genCorrRedBHi}{-0.135}
\newcommand{\genCorrRedBP}{3.8\times10^{-6}}
\newcommand{\genCorrRedBRes}{+0.010}
\newcommand{\genCorrTargetZZ}{-0.400}
\newcommand{\genDMn}{100}

\newcommand{\genDMpopDev}{7.5\times10^{-16}}
\newcommand{\genDMherm}{5.7\times10^{-17}}
\newcommand{\genDMtrace}{3.8\times10^{-15}}
\newcommand{\genDMminev}{-2.1\times10^{-16}}
\newcommand{\genDMrankThree}{59}
\newcommand{\genDMrankFour}{41}
\newcommand{\genDMrankOther}{0}
\newcommand{\genDMsimRank}{1}
\newcommand{\genDMcohTV}{0.048}
\newcommand{\genDMcohDiff}{0.097}
\newcommand{\genDMstateF}{0.438}
\newcommand{\genDMstateFlo}{0.220}
\newcommand{\genDMstateFhi}{0.683}
\newcommand{\genDMclF}{0.897}
\newcommand{\genDMpurity}{0.430}
\newcommand{\genDMspearman}{0.55}
\newcommand{\genDMpCtl}{0.561}
\newcommand{\genDMCtlA}{0.384}
\newcommand{\genDMCtlB}{0.379}
\newcommand{\genDMpRed}{0.007}
\newcommand{\genDMRedA}{0.481}
\newcommand{\genDMRedB}{0.527}
\newcommand{\genSessOrder}{D3, D2, D5, D6, D4, D1, D0}
\newcommand{\genSessNtotal}{700}
\newcommand{\genSessHours}{24.3}
\newcommand{\genSessSpanLo}{3.4}
\newcommand{\genSessSpanHi}{3.5}
\newcommand{\genSessFirst}{31~Mar}
\newcommand{\genSessLast}{2~Apr}
\newcommand{\genSpamQone}{0.073}
\newcommand{\genSpamQtwo}{0.058}
\newcommand{\genSpamQtwoLone}{0.128}
\newcommand{\genSpamKWqone}{1.7\times10^{-6}}
\newcommand{\genSpamKWqtwo}{7.9\times10^{-9}}
\newcommand{\genSpamKWLone}{1.6\times10^{-22}}
\newcommand{\genSpamRangeLo}{0.048}
\newcommand{\genSpamRangeHi}{0.075}
\newcommand{\genSpamRangeRatio}{1.6}
\newcommand{\genSpamFidRho}{+0.21}
\newcommand{\genSpamFidP}{0.64}
\newcommand{\genSpamOrderRho}{-0.04}
\newcommand{\genSpamOrderP}{0.94}
\newcommand{\genLeakLzeroLo}{0.079}
\newcommand{\genLeakLzeroHi}{0.129}
\newcommand{\genLeakLoneLo}{0.095}
\newcommand{\genLeakLoneHi}{0.163}
\newcommand{\genFidGapLo}{+0.0034}
\newcommand{\genFidGapHi}{+0.0232}
\newcommand{\genFidGapMed}{+0.0094}
\newcommand{\genFidGapNcells}{14}
\newcommand{\genTVidentityMax}{1.2\times10^{-10}}
\newcommand{\genTVidentityN}{350}
\newcommand{\genRoCond}{1.37}
\newcommand{\genRoDiag}{0.855}
\newcommand{\genRoRawFid}{0.927}

\newcommand{\genRoMitFid}{0.922}

\newcommand{\genRoMixFid}{0.926}

\newcommand{\genRootAngleLo}{45.57}
\newcommand{\genRootAngleHi}{106.22}
\newcommand{\genRootAngleN}{4}
\newcommand{\genExactMax}{4.4\times10^{-16}}
\newcommand{\genExactAB}{1.1\times10^{-16}}
\newcommand{\genExactNcirc}{14}
\newcommand{\genGatesLzero}{3}
\newcommand{\genGatesLzeroNoTail}{1}
\newcommand{\genGatesLone}{13}
\newcommand{\genGatesLoneNoTail}{11}
\newcommand{\genGatesFull}{21}
\newcommand{\genGatesFullNoTail}{19}
\newcommand{\genAngleRms}{3.73}
\newcommand{\genAngleRmsLo}{3.42}
\newcommand{\genAngleRmsHi}{4.05}
\newcommand{\genAngleWithin}{1.65}
\newcommand{\genAngleBetween}{3.54}
\newcommand{\genAnglePredicted}{0.33}
\newcommand{\genAngleKW}{1.6\times10^{-23}}
\newcommand{\genAngleMeanLo}{-4.65}
\newcommand{\genAngleMeanHi}{+4.91}

\newcommand{\genDescDeltaRho}{-0.64}
\newcommand{\genDescDeltaP}{0.12}
\newcommand{\genDescEntRho}{+0.39}
\newcommand{\genDescContRho}{-0.11}
\newcommand{\genDescCollinear}{0.93}
\newcommand{\genDescCollinearP}{0.003}
\newcommand{\genMatchSpan}{0.014}
\newcommand{\genMatchList}{D6, D2, D5, D4}
\newcommand{\genMatchFidLo}{0.882}
\newcommand{\genMatchFidHi}{0.964}
\newcommand{\genStagePredMaxRho}{+0.71}
\newcommand{\genStagePredMaxP}{0.071}

\newcommand{\genStageMono}{4}
\newcommand{\genStageBelowList}{D5}
\newcommand{\genSchedSpan}{3.72}
\newcommand{\genSchedMedianGap}{2.1}
\newcommand{\genSchedPosLo}{50.4}
\newcommand{\genSchedPosHi}{50.6}
\newcommand{\genSchedDriftRho}{+0.13}
\newcommand{\genSchedDriftP}{0.20}
\newcommand{\genSchedDriftSlope}{3.0\times10^{-4}}
\newcommand{\genSchedGapMin}{17.4}
\newcommand{\genSchedGapAt}{54}
\newcommand{\genSchedGapBefore}{0.8832}
\newcommand{\genSchedGapAfter}{0.8962}
\newcommand{\genSchedGapP}{0.37}
\newcommand{\genMultFlipTarget}{D3}
\newcommand{\genMultFlipRaw}{0.0145}
\newcommand{\genMultFlipRank}{4}
\newcommand{\genMultFlipHolmFactor}{4}
\newcommand{\genMultFlipBHFactor}{1.75}
\newcommand{\genMultFlipHolm}{0.0580}
\newcommand{\genMultFlipBH}{0.0254}
\newcommand{\genLzeroSubsetTargets}{D0, D1, D3, D4}
\newcommand{\genLzeroSubsetN}{4}
\newcommand{\genLzeroSubsetRoot}{90}
\newcommand{\genLzeroSubsetKW}{5.9\times10^{-5}}
\newcommand{\genLzeroSubsetLo}{0.0475}
\newcommand{\genLzeroSubsetHi}{0.0729}
\newcommand{\genLzeroSubsetRatio}{1.53}
\newcommand{\genExportNjobs}{1239}
\newcommand{\genExportFrom}{30~January~2026}
\newcommand{\genExportTo}{7~October~2026}
\newcommand{\genDMrecN}{948}

\newcommand{\genDMrecSpan}{March~2026 to October~2026}

\newcommand{\genDMrecDiagN}{948}
\newcommand{\genDMrecRanks}{2, 3, 4, 5}
\newcommand{\genDMrecRankHist}{rank 2: 49; rank 3: 411; rank 4: 464; rank 5: 24}
\newcommand{\genDMrecZeroMax}{4.3\times10^{-16}}
\newcommand{\genDMrecZeroMedian}{2.3\times10^{-17}}
\newcommand{\genDMrecNonzeroMin}{5.8\times10^{-5}}
\newcommand{\genDMrecGap}{11}
\newcommand{\genDMrecMid}{13}
\newcommand{\genDMrecTotalEv}{7584}
\newcommand{\genAngN}{25}

\newcommand{\genAngInterEst}{-0.0296}
\newcommand{\genAngInterLo}{-0.0606}
\newcommand{\genAngInterHi}{+0.0014}
\newcommand{\genAngInterP}{0.061}

\newcommand{\genAngCmdFolded}{686.7890}

\providecommand{\mref}[1]{\ref{#1}}
\providecommand{\supref}[1]{\ref{#1}}

\usepackage{hyperref}\usepackage{url}\usepackage{xurl}
\usepackage{array,longtable,float}\usepackage{subcaption}
\usepackage{tikz}\usepackage{pgfplots}\pgfplotsset{compat=1.18}
\usepackage[normalem]{ulem}
\usetikzlibrary{positioning,arrows.meta,shapes.geometric,calc,decorations.pathreplacing,patterns}
\usepackage{quantikz}
\renewcommand{\ket}[1]{|#1\rangle}
\providecommand{\dg}{\ensuremath{{}^{\circ}}}

\begin{document}

\title{Experimental characterisation of compilation-dependent
Grover--Rudolph probability distribution loading on a three-qubit NMR
processor}

\author{Antonio Falc\'o$^{1}$ \and Daniela Falc\'o-Pomares$^{2}$ \and
        Juan-Carlos Latorre$^{1}$ \and Yuefeng Lin$^{3}$}
\titlerunning{Compilation-dependent probability distribution loading}
\authorrunning{A. Falc\'o et al.}
\institute{Antonio Falc\'o \at
  Departamento de Matem\'aticas, F\'{\i}sica y Ciencias Tecnol\'ogicas,
  Universidad CEU Cardenal Herrera, CEU Universities,
  San Bartolom\'e 55, 46115 Alfara del Patriarca (Valencia), Spain \\
  \email{antonio.falco@uchceu.es}
\and
  Daniela Falc\'o-Pomares \at
  Grupo de Investigaci\'on Bisite, Universidad de Salamanca,
  Calle Espejo s/n, 37007 Salamanca, Spain
\and
  Juan-Carlos Latorre \at
  Departamento de Matem\'aticas, F\'{\i}sica y Ciencias Tecnol\'ogicas,
  Universidad CEU Cardenal Herrera, CEU Universities,
  San Bartolom\'e 55, 46115 Alfara del Patriarca (Valencia), Spain
\and
  Yuefeng Lin \at
  Shenzhen SpinQ Technology Co., Ltd, Shenzhen, 518048, China}
\date{}
\maketitle

\ifhassvjourthree\else
\begin{center}\small
$^{1}$Departamento de Matem\'aticas, F\'{\i}sica y Ciencias Tecnol\'ogicas,
Universidad CEU Cardenal Herrera, CEU Universities,
San Bartolom\'e 55, 46115 Alfara del Patriarca (Valencia), Spain\\
$^{2}$Grupo de Investigaci\'on Bisite, Universidad de Salamanca,
Calle Espejo s/n, 37007 Salamanca, Spain\\
$^{3}$Shenzhen SpinQ Technology Co., Ltd, Shenzhen, 518048, China\\[3pt]
Corresponding author: \email{antonio.falco@uchceu.es}
\end{center}
\medskip
\fi

\begin{abstract}
Exact state-preparation circuits are normally judged by how closely the measured
output distribution reproduces the target. We use one, the ancilla-free
Grover--Rudolph construction compiled into $\{\Ry,X,\mathrm{CNOT}\}$, to ask
instead how much the accuracy of a probability-distribution loading depends on
which of several logically equivalent compilations is executed.

Two Gray-code realisations of the level-2 uniformly controlled rotation, A and
B, differ only in the order of their CNOT controls: same unitary, same angles.
On a three-qubit liquid-state NMR processor they do not perform identically.
Across $700$ runs over seven targets the A/B contrast is significant for
\genNsigHolm{} of them under Holm, with no fixed sign.

Two further campaigns isolate the effect within one unit. Removing the
circuit's identity-equivalent gates, which changes the executed schedule but not
the unitary, affects the two realisations differently: a
ladder~$\times$~intervention interaction of $\genInterEst$ (95\% CI
$[\genInterLo,\genInterHi]$, paired $p=\genInterP$), reversing the sign of the
A/B advantage. Across five acquisitions of this one logical circuit the contrast
runs from $\genCmpCtlAB$ to $\genCmpRedAB$. Not every compilation choice
matters, however: commanding the two level-1 inverse rotations as
$720\dg-\theta$ rather than $-\theta$, with the gate list unchanged, moves the
mean fidelity by $-0.0006$ ($[-0.0157,+0.0146]$) on one target in one session.
We offer the order and timing of the executed pulses as the hypothesis these
observations favour, without identifying a mechanism.

The metric does not certify coherences, and the measured correlators are
inconsistent with any global depolarising description: on the uniform target, at
a classical fidelity of $\genCorrDzeroF$, a two-body correlation of
$\genCorrDzeroZZ$ appears where the target has none. Run-level data and code are deposited for every
campaign run on our own unit; the pulse-level campaign on the manufacturer's
unit is reported from summary tables, as stated there.
\end{abstract}


\section{The loading problem and an exact reference}
\label{sec:target}

A state-preparation routine is specified by the unitary it implements. This
paper asks what that specification leaves undetermined once the circuit reaches
hardware, and measures what it costs.

The vehicle is the Grover--Rudolph construction
\cite{Grover2000,GroverRudolph2002}, the canonical exact method for loading a
classical probability distribution into the amplitudes of a register
\cite{Brassard2002,Montanaro2015,Zoufal2019,Herbert2021}. Compiled ancilla-free
into $\{\Ry,X,\mathrm{CNOT}\}$, it leaves choices the ideal model cannot
distinguish because they produce the same operator: which Gray-code order the
final uniformly controlled rotation traverses, whether gates that act as the
identity are emitted at all, which representative of an angle is commanded. We
hold the algorithm, the target and the gate set fixed, vary only those choices,
and measure the output.

An exact construction is what makes the question answerable. The ideal output
is known in closed form, so the measured law can be compared with the target
rather than with a simulator, and every discrepancy is implementation with no
approximation error to separate out first. This is the regime in which the
compilation layer and the physical layer cannot be assessed independently
\cite{Preskill2018}. Three qubits keeps the circuit short
enough that one compilation decision can be isolated and intervened upon.

Four campaigns on two SpinQ Triangulum processors follow. The first
(Section~\ref{sec:suite}), $700$ runs over seven targets, establishes that two Gray-code realisations of the
same unitary do not perform alike and that the difference has no fixed sign.
The second (Section~\ref{sec:intervention}) removes gates whose joint action is
the identity --- changing the executed schedule, not the operator --- and finds
that the two realisations respond to that change differently, by enough to
reverse which of them is better. The third
(Section~\ref{subsec:long-angle}) commands the two level-1 inverse rotations as
$720\dg-\theta$ instead of $-\theta$, with the gate list untouched, and finds
no detectable change in mean fidelity. The
fourth (Section~\ref{sec:pulse}), on the manufacturer's unit, intervenes on the
pulse layer itself. Taken together they say that the sensitivity is not a
property of the target, and they make the order and timing of the executed
pulses the hypothesis to beat --- without establishing it, and without ruling
out the alternatives.

We do not identify a mechanism, and Section~\ref{sec:discussion} is explicit
about which statements are observations, which are hypotheses, and which
questions the designs used here cannot settle.

The paper is arranged so that each claim can be checked against the design that
produced it. Section~\ref{sec:design} sets out the processor and its
calibration regime, the compiled circuits as the instrument executed them, the
acquisition schedules, the measurement conventions and the metrics, and states
for each campaign what it can and cannot separate. Sections~\ref{sec:suite}
to~\ref{sec:pulse} report the four campaigns in the order above.
Section~\ref{sec:density-record} turns to what the output distributions leave
undetermined about the prepared state, and why the state-level information in
the instrument record cannot yet settle it. Section~\ref{sec:discussion}
separates what was observed from what is hypothesis and from what remains open,
and lists the experiments that would close the gap. A supplement carries the
construction and its compilation, the truncated-stage analysis, the
multiplicity treatment, the extended statistics of each campaign and the full
protocol; the main text references it by section throughout, and never repeats
it.

\subsection{Related work}
\label{subsec:related}

Recursive splitting of probability mass originates with Grover
\cite{Grover2000} and the efficiently-integrable formulation of Grover and
Rudolph \cite{GroverRudolph2002}, with related constructions by Kaye and Mosca
\cite{KayeMosca2001}. The synthesis cost was settled by M\"ott\"onen \emph{et
al.}, who introduced uniformly controlled rotations and the Gray-code ladder
\cite{Mottonen2005,Bergholm2005}, and by Shende, Bullock and Markov
\cite{ShendeBullockMarkov2006,NielsenChuang2010}; later work trades depth against ancillas
\cite{PleschBrukner2011,Araujo2021,ZhangLiYuan2022}, or replaces the exact
construction by an approximate one, as the Walsh-series method of Zylberman and
Debbasch does \cite{ZylbermanDebbasch2024}. That literature answers how many
gates a loading costs. A separate line questions whether
Grover--Rudolph loading survives an end-to-end complexity accounting
\cite{Herbert2021,Montanaro2015,Sanders2019,Bravyi2021}; both concern the
circuit as an object in the ideal model, and neither addresses what varying its
compilation does on hardware. Liquid-state NMR remains a well-understood
platform for small-scale control \cite{VandersypenChuang2005,Khaneja2005} and
the Triangulum is documented in \cite{FengEtAl2022}; characterisation work on
such devices reports gate and readout fidelities rather than the sensitivity of
a fixed algorithmic circuit to compilation.

\subsection{The construction and its compilation}
\label{subsec:construction}

For a target law $p$ on $\{0,1\}^n$, write $p_w$ for the mass of the dyadic node
reached by the prefix $w$. The Grover--Rudolph splitting angle at that node is
defined by
\begin{equation}
\cos^{2}\theta_w=\frac{p_{w0}}{p_w},\qquad
\sin^{2}\theta_w=\frac{p_{w1}}{p_w},
\qquad \theta_w\in[0,\pi/2],
\label{eq:angle-map}
\end{equation}
and the gate the hardware executes is $\Ry(\phi_w)$ with $\phi_w=2\theta_w$. We
keep the two symbols distinct throughout: $\theta_w$ is the splitting angle of
the construction, $\phi_w$ the commanded rotation of the compiled circuit.

The consequence that matters here is structural. Refining level $k$ of the tree
applies, to the active qubit, a rotation whose angle is conditioned on the
prefix already prepared --- that is, a uniformly controlled $\Ry$ with $k$
controls. Such an operation admits an ancilla-free decomposition into
$\{\Ry(\cdot),X,\mathrm{CNOT}\}$ by the Gray-code ladder of
\cite{Mottonen2005}, at a cost of $2^{k}$ rotations and $2^{k}$ CNOTs. The
ladder visits the $2^{k}$ control patterns in a Gray-code order, and that order
is not determined by the construction. For $n=3$ the final level has two
controls, four slots, and two natural traversals, which are the realisations
compared throughout this paper and are drawn in
Figure~\ref{fig:circuits}. Avoiding ancillas is not only a matter of elegance on
a three-qubit device: it is what makes the whole circuit fit.

The circuit is organised by level, and two truncations are acquired alongside
the complete one: $L0$ applies the root rotation alone, $L01$ applies levels
$0$ and $1$, and FULL applies all three. With the identity gates the compiler
emits they carry $\genGatesLzero$, $\genGatesLone$ and $\genGatesFull$ gates
respectively; without them, $\genGatesLzeroNoTail$, $\genGatesLoneNoTail$ and
$\genGatesFullNoTail$. Section~\ref{subsec:stages} states what the truncated
stages are used for, which is less than they might appear to offer.

The rigorous dyadic formulation, the proof that the construction reproduces the
target exactly, and the bounds used in the error discussion are given in the
companion paper \cite{FalcoFalcoPomaresMatthies2026} and are not reproduced
here. Supplement~\supref{sup:theory} collects the notation and the specific
results this paper depends on; it is a dependency list, not a summary.

\subsection{The benchmark targets}
\label{subsec:targets}

\begin{table}[htbp]
\caption{The seven benchmark laws. $H$ is the Shannon entropy in bits, the
contrast is the ratio of largest to smallest mass, and the last column is the
total-variation distance from the uniform law. The suite was chosen to separate
these descriptors from one another: D2 and D5 have similar distance from uniform
and entropies within $0.09$ bits, but contrasts differing by a factor of four.}
\label{tab:targets}
\centering\small
\begin{tabular}{llccc}
\toprule
Target & Law & $H$ (bits) & $p_{\max}/p_{\min}$ & $\TV$ to uniform \\
\midrule
D0 & $(0.125,\,0.125,\,0.125,\,0.125,\,0.125,\,0.125,\,0.125,\,0.125)$ & $3.00$ & $1$ & $0.000$ \\
D1 & $(0.05,\,0.1,\,0.15,\,0.2,\,0.2,\,0.15,\,0.1,\,0.05)$ & $2.85$ & $4$ & $0.200$ \\
D2 & $(0.35,\,0.25,\,0.15,\,0.1,\,0.07,\,0.04,\,0.03,\,0.01)$ & $2.45$ & $35$ & $0.375$ \\
D3 & $(0.02,\,0.03,\,0.05,\,0.4,\,0.4,\,0.05,\,0.03,\,0.02)$ & $2.02$ & $20$ & $0.550$ \\
D4 & $(0.02,\,0.05,\,0.12,\,0.31,\,0.31,\,0.12,\,0.05,\,0.02)$ & $2.44$ & $16$ & $0.370$ \\
D5 & $(0.214,\,0.34,\,0.002,\,0.201,\,0.004,\,0.037,\,0.119,\,0.083)$ & $2.36$ & $144$ & $0.380$ \\
D6 & $(0.052,\,0.023,\,0.17,\,0.116,\,0.091,\,0.06,\,0.446,\,0.042)$ & $2.41$ & $19$ & $0.366$ \\
\bottomrule
\end{tabular}

\end{table}

The seven targets of Table~\ref{tab:targets} were chosen so that the obvious
descriptors of a distribution do not move together. Entropy ranges over
$2.02$ to $3.00$ bits, contrast over a factor of $144$, and distance from the
uniform law from $0$ to $0.55$, and these three orderings disagree. D0 is the
uniform law and serves as a control: at the final level all four of its
sub-rotations are equal, so the two ladders issue identical commands and differ
only in the order of their CNOT controls, which isolates scheduling from angle
content (Section~\ref{subsec:suite-d0}). D5 has two masses below $0.005$ ($0.002$ and
$0.004$), whereas the smallest mass in D6 is $0.023$. The low masses in D5
bring sub-rotations close to zero and exercise the ladder's degenerate slots.

One property of the suite is worth flagging here because it is easy to overlook
and does real work later: four of these seven laws have all three single-qubit
marginals exactly equal to $1/2$. For those targets a report of marginal balance
is uninformative by construction, and everything that distinguishes the target
from the uniform law lives in its correlations. Section~\ref{subsec:correlators}
takes that up.

\subsection{Exactness in the ideal model}
\label{subsec:exactness}

The comparison this paper rests on has content only if the two realisations are
genuinely equivalent before the hardware sees them. They are. Evaluated
analytically, the compiled FULL-stage circuit reproduces its target to
$\genExactMax$ in the worst case over all $\genExactNcirc$ circuits of the suite
--- seven targets, two ladders --- and the two ladders agree with each other to
$\genExactAB$. Both figures are at the level of double-precision arithmetic, and
the check is part of the deposited verification suite, which needs neither
hardware nor the vendor SDK.

So in the ideal model the A/B comparison has no content whatever: the two
circuits implement the same unitary, produce the same state, and yield the same
distribution. Any difference measured on hardware is a property of the
implementation and not of the algorithm. That is the whole basis of what
follows, and it is also why the result reported in
Section~\ref{sec:compilations} --- where a third compilation of the same logical
circuit, equally exact, gives a different answer again --- is a statement about
devices rather than about mathematics.
\section{Experimental design and measurement}
\label{sec:design}

\subsection{The processor and its calibration state}
\label{subsec:platform}

The hardware is the SpinQ Triangulum, a three-qubit room-temperature
liquid-state NMR processor \cite{FengEtAl2022}. Two units of this model were
used, and only two exist in this study. All gate-level work reported here --- the
benchmark campaign of Section~\ref{sec:suite}, the intervention campaign of
Section~\ref{sec:intervention}, the compilation comparison of
Section~\ref{sec:compilations} and the readout session below --- was carried out
on the unit at Universidad CEU Cardenal Herrera. The pulse-level campaign of
Section~\ref{sec:pulse} was carried out on the unit at Shenzhen SpinQ
Technology. Absolute fidelities from different units are not
interchangeable; where this paper compares across units it compares effects,
not absolute numbers.

Calibration of these devices is performed on demand rather than on a schedule:
the operator recalibrates when a loss of fidelity is observed. Two consequences
follow. Calibration state is a property of a unit at a time, so absolute
fidelities are comparable only within a session on one unit; and because
recalibration is triggered by observed performance rather than by elapsed time,
calibration events are correlated with the quantity being measured. The devices
retain no calibration log, so the events in any interval are neither dated nor
counted. Section~\ref{subsec:stages} gives the one observable in this design
that measures the instrument's state without reference to the target.

\subsection{The compiled circuits}
\label{subsec:circuits}

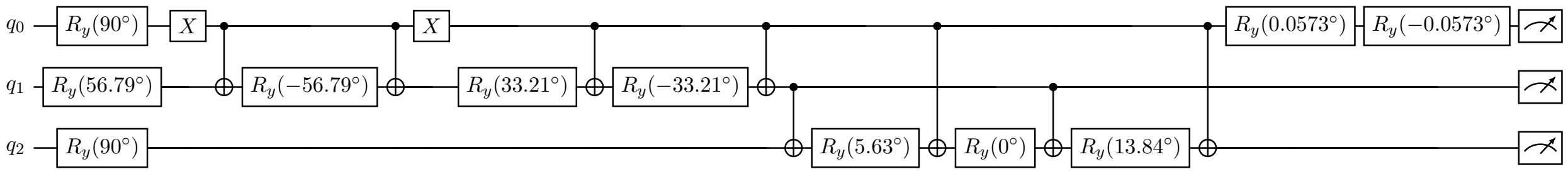
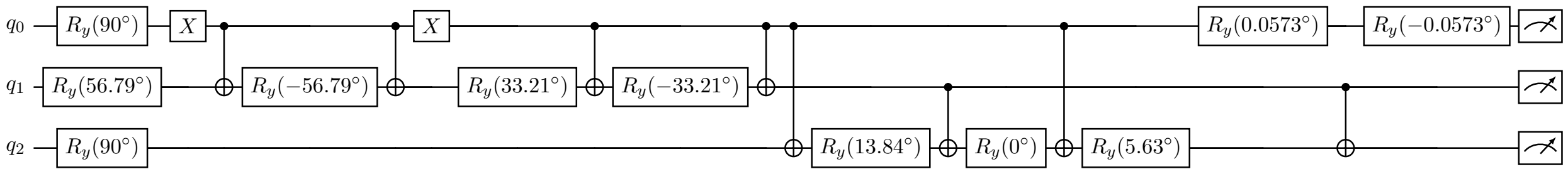
\begin{figure}[htbp]
\centering
\begin{subfigure}[b]{\linewidth}
\centering
\resizebox{\linewidth}{!}{%
\begin{quantikz}[row sep=0.40cm, column sep=0.16cm]
\lstick{$q_0$} &
  \gate{\Ry(90\dg)} & \gate{X} & \ctrl{1} & \qw & \ctrl{1} & \gate{X} &
  \qw & \ctrl{1} & \qw & \ctrl{1} &
  \qw & \qw & \ctrl{2} & \qw & \qw & \qw & \ctrl{2} &
  \gate{\Ry(0.0573\dg)} & \gate{\Ry(-0.0573\dg)} & \meter{} \\
\lstick{$q_1$} &
  \gate{\Ry(56.79\dg)} & \qw & \targ{} & \gate{\Ry(-56.79\dg)} & \targ{} & \qw &
  \gate{\Ry(33.21\dg)} & \targ{} & \gate{\Ry(-33.21\dg)} & \targ{} &
  \ctrl{1} & \qw & \qw & \qw & \ctrl{1} & \qw & \qw & \qw & \qw & \meter{} \\
\lstick{$q_2$} &
  \gate{\Ry(90\dg)} & \qw & \qw & \qw & \qw & \qw & \qw & \qw & \qw & \qw &
  \targ{} & \gate{\Ry(5.63\dg)} & \targ{} & \gate{\Ry(0\dg)} & \targ{} &
  \gate{\Ry(13.84\dg)} & \targ{} & \qw & \qw & \meter{}
\end{quantikz}}
\caption{Ladder~A. Level-2 control sequence $q_1,q_0,q_1,q_0$.}
\label{fig:circ-A}
\end{subfigure}

\vspace{0.4cm}

\begin{subfigure}[b]{\linewidth}
\centering
\resizebox{\linewidth}{!}{%
\begin{quantikz}[row sep=0.40cm, column sep=0.16cm]
\lstick{$q_0$} &
  \gate{\Ry(90\dg)} & \gate{X} & \ctrl{1} & \qw & \ctrl{1} & \gate{X} &
  \qw & \ctrl{1} & \qw & \ctrl{1} &
  \ctrl{2} & \qw & \qw & \qw & \ctrl{2} & \qw &
  \gate{\Ry(0.0573\dg)} & \qw & \gate{\Ry(-0.0573\dg)} & \meter{} \\
\lstick{$q_1$} &
  \gate{\Ry(56.79\dg)} & \qw & \targ{} & \gate{\Ry(-56.79\dg)} & \targ{} & \qw &
  \gate{\Ry(33.21\dg)} & \targ{} & \gate{\Ry(-33.21\dg)} & \targ{} &
  \qw & \qw & \ctrl{1} & \qw & \qw & \qw & \qw & \ctrl{1} & \qw & \meter{} \\
\lstick{$q_2$} &
  \gate{\Ry(90\dg)} & \qw & \qw & \qw & \qw & \qw & \qw & \qw & \qw & \qw &
  \targ{} & \gate{\Ry(13.84\dg)} & \targ{} & \gate{\Ry(0\dg)} & \targ{} &
  \gate{\Ry(5.63\dg)} & \qw & \targ{} & \qw & \meter{}
\end{quantikz}}
\caption{Ladder~B. Level-2 control sequence $q_0,q_1,q_0,q_1$, which exchanges
the two non-zero sub-rotations and splits the tail across the final CNOT.}
\label{fig:circ-B}
\end{subfigure}

\caption{The compiled FULL-stage circuits for~D1, transcribed from the gate
lists the instrument recorded for the October campaign. Angles are commanded
values in degrees. The two panels have identical gate counts and identical
angles, and differ only in the Gray-code ordering of the level-2 block on $q_2$.
Two gates act as the identity in the ideal model and are nevertheless executed:
the zero-angle rotation inside the level-2 ladder, and the tail pair
$\Ry(+0.0573\dg)\,\Ry(-0.0573\dg)$ on $q_0$, whose two pulses are adjacent in
ladder~A but separated by the final CNOT in ladder~B.
Section~\ref{sec:intervention} is defined by removing both.}
\label{fig:circuits}
\end{figure}

The compiled FULL-stage circuit in the native gate set
$\{\Ry(\cdot),X,\mathrm{CNOT}\}$ is shown in Figure~\ref{fig:circuits}. It
carries $21$ gates: $11$ rotations, $8$ CNOTs and $2$ $X$ gates. The level-1
uniformly controlled rotation on $q_1$ is realised as
$\Ry(+\theta/2)\,\mathrm{CNOT}\,\Ry(-\theta/2)\,\mathrm{CNOT}$, bracketed by $X$
gates on $q_0$ for the branch conditioned on $q_0=0$. The level-2 rotation on
$q_2$ is realised as the four-step Gray-code ladder whose two orderings define
the comparison this paper is about: ladder~A applies controls
$q_1,q_0,q_1,q_0$ and ladder~B applies $q_0,q_1,q_0,q_1$, which exchanges the
two non-zero sub-rotations. Everything else in the two circuits is identical,
gate for gate and angle for angle.

Two of the $21$ gates implement the identity: a zero-angle rotation inside the
level-2 ladder, emitted because the ladder has four slots and this target
populates only three, and the tail pair on $q_0$, delivered as two physical
pulses. Section~\ref{sec:intervention} measures what happens when both are
removed.

The representative in which a negative angle is commanded determines the pulse
the instrument delivers, so it is not a matter of presentation. The figure shows
the October compilation; the compiler that produced the earlier campaign used a
different representative, and Section~\ref{subsec:long-three} sets out the
difference.

Two truncated stages were acquired alongside the full circuit and are used only
as controls. $L0$ retains the level-0 rotation on $q_0$ alone; $L01$ retains
levels $0$ and $1$. Their compiled circuits are in
Supplement~\supref{sup:circuits}, and Section~\ref{subsec:stages} states what
they can and cannot be asked.

\subsection{Acquisition}
\label{subsec:acquisition}

The benchmark campaign comprises $\genSessNtotal$ hardware runs over seven
targets. Each target contributes $\genNreps$ runs in each of four cells: FULL
with ladder~A, FULL with ladder~B, $L0$ and $L01$. The truncated stages were
acquired with ladder~A only, so they contribute nothing to any A/B comparison
and are not used in one.

\begin{table}[htbp]
\caption{Acquisition structure of the benchmark campaign, reconstructed from the
run-level timestamps. Each target occupies one contiguous session; the sessions
are consecutive and no target is split across two of them.}
\label{tab:sessions}
\centering\small
\begin{tabular}{llccc}
\toprule
Order & Target & Session start (UTC) & Span & Runs \\
\midrule
1 & D3 & 31~Mar, 18:01 & $3.4$\,h & 100 \\
2 & D2 & 31~Mar, 21:42 & $3.4$\,h & 100 \\
3 & D5 & 1~Apr, 07:11 & $3.5$\,h & 100 \\
4 & D6 & 1~Apr, 12:07 & $3.5$\,h & 100 \\
5 & D4 & 1~Apr, 15:37 & $3.5$\,h & 100 \\
6 & D1 & 1~Apr, 20:30 & $3.5$\,h & 100 \\
7 & D0 & 2~Apr, 07:17 & $3.5$\,h & 100 \\
\bottomrule
\end{tabular}

\end{table}

Table~\ref{tab:sessions} gives the structure. One target was acquired per
session, each session running $\genSessSpanLo$ to $\genSessSpanHi$ hours, in the
order $\genSessOrder$, between $\genSessFirst$ and $\genSessLast$ 2026, for
$\genSessHours$ hours of acquisition in total. This one-target-per-session
design is the aliasing that Section~\ref{subsec:suite-limits} is about: nothing
in this campaign separates a property of a distribution from a property of the
hours in which it was measured. Within a session the FULL-stage runs were
acquired in alternating blocks of five rather than interleaved, so ladder~B sits
a mean of five slots later than ladder~A; the consequences are quantified in
Section~\ref{subsec:long-caveats}.

The intervention campaign of Section~\ref{sec:intervention} was acquired six
months later on the same unit, in a single session, with the cells interleaved
run by run and paired by repetition. Its design is given there.

\paragraph{What a ``shot'' is on this instrument.} The SDK accepts a
\texttt{shots} parameter and the run-level log records it --- $2048$ for the
October runs, $4096$ as configured for the benchmark campaign, which does not
record the value, and nothing at all in the instrument's job database. The
parameter does not mean what the name suggests. The manufacturer confirms that
Triangulum reads out an NMR ensemble, and that in the software path inspected
the nominal \texttt{shots} value scales the measured probabilities into
nominal counts rather than counting independent projective measurements.

Two consequences run through the paper. There is no multinomial sampling scale
attached to a run, so a difference between $2048$ and $4096$ is not a
difference in statistical precision and we draw no inference from it. And the
only quantity that does carry sampling variation is the spread across repeated
hardware executions, which is why every uncertainty reported here is computed
across runs, and why the designs that matter are paired within a session.

\subsection{Measurement conventions}
\label{subsec:conventions}

\paragraph{Bit order.} The mapping between the instrument's readout channels and
the bit positions of a computational basis label was measured on the unit rather
than assumed. An $X$ gate was applied to each qubit in turn with the rest left
idle, and the dominant outcomes were \texttt{100}, \texttt{010} and \texttt{001}
respectively, confirming the canonical most-significant-first convention. The
dominant probabilities were $0.657$, $0.805$ and $0.864$: the readout is visibly
imperfect, and most so on $q_0$, but the assignment is unambiguous. This check
was run fifteen minutes before the October campaign on the same unit, and its
jobs are in the instrument export with the campaign's own.

\paragraph{Raw distributions.} All distributions reported in this paper are
raw. No readout mitigation is applied anywhere, for an experimental reason.

\begin{table}[htbp]
\caption{Effect of confusion-matrix readout mitigation at the FULL stage,
measured against the target law. From a separate session on the same unit (D1,
ladder~B, five runs per stage, at a nominal $2048$ shots), independent of the benchmark
campaign. The partial correction mixes the raw and inverted distributions with
weight $\lambda=0.3$.}
\label{tab:readout}
\centering\small
\begin{tabular}{lccc}
\toprule
Post-processing & $\TV$ & $\ell^2$ & Classical fidelity \\
\midrule
Raw (no mitigation) & $0.245$ & $0.192$ & $0.927$ \\
Mitigated (confusion-matrix inverse) & $0.251$ & $0.203$ & $0.922$ \\
Partially mitigated ($\lambda=0.3$) & $0.247$ & $0.195$ & $0.926$ \\
\bottomrule
\end{tabular}

\end{table}

A full $8\times8$ confusion matrix was estimated on this unit by preparing each
computational basis state in turn and recording the measured output. The matrix
is well conditioned --- condition number $\genRoCond$, mean diagonal
$\genRoDiag$ --- so inverting it is numerically harmless. Applying it to the
FULL-stage output nevertheless leaves the result slightly \emph{worse}:
Table~\ref{tab:readout} reports a classical fidelity of $\genRoRawFid$ raw
against $\genRoMitFid$ mitigated and $\genRoMixFid$ partially mitigated, with
$\TV$ moving the same way in all three. Since the matrix is well conditioned this cannot be
laid to an ill-posed inversion.

What that establishes is that this correction, estimated this way, did not
improve the FULL-stage result in this session. It does not establish that the
residual is not readout-dominated: conditioning speaks to the stability of the
inversion, not to whether the estimated matrix represents the readout process,
and the calibration states are themselves prepared with error, the matrix
carries an uncertainty we do not propagate, and its transferability across
sessions is untested. The reading we find plausible --- that the register is
already distorted by circuit-level errors, so removing the measurement bias
re-weights a distribution whose shape is wrong for other reasons --- is
consistent with the observation and not demonstrated by it.

That observation rests on this experiment and on nothing else. It is in
particular not inferred from the agreement between the distance to the ideal
simulator and the distance to the target: at the FULL stage the simulator
\emph{is} the target, so the two coincide by construction --- to
$\genTVidentityMax$ over all $\genTVidentityN$ FULL runs --- and carry no
information about readout.

\subsection{Metrics and what they certify}
\label{subsec:metrics}

A run yields a probability vector over the eight computational basis outcomes.
Every quantity this paper reports is a functional of that vector: the total
variation distance $\TV$, the $\ell^2$ distance, the classical fidelity
$F(p,q)=\bigl(\sum_k\sqrt{p_kq_k}\bigr)^2$, the three single-qubit marginals,
and the two- and three-body correlators of
Section~\ref{subsec:correlators}. None of them certifies a coherence, and none
of them distinguishes the prepared state from any other state with the same
measurement statistics in this basis. Section~\ref{sec:density-record} returns
to what would be needed to go further. They also differ sharply in sensitivity;
Section~\ref{subsec:correlators} opens with how lenient the classical fidelity
is on this suite.

The mean of the per-run fidelities and the fidelity of the mean measured law
are different numbers, the second the larger by concavity, and over the
$\genFidGapNcells$ FULL-stage cells the gap runs from $\genFidGapLo$ to
$\genFidGapHi$ with median $\genFidGapMed$ --- the same order as some of the
contrasts reported here. We use the mean of per-run values throughout, so that
the run-to-run spread is visible and the tests operate on runs rather than on a
pooled law. Section~\ref{sec:pulse} uses the same convention: its D1 table
reports the mean and sample standard deviation of five per-execution
fidelities in each configuration--ladder cell, and its direct-pulse benchmark
table pools 50 per-execution values per target. Distances between mean output
distributions are identified separately and are not means of per-run
distances. Comparisons between the Shenzhen and CEU campaigns remain
confounded by the device and acquisition period.

\subsection{Truncated stages as auxiliary controls}
\label{subsec:stages}

The three stages are not a depth series and we do not use them as one. Moving
from $L0$ to $L01$ to FULL changes the gate count, the ideal state and the
support of the ideal distribution simultaneously, so no comparison among them
isolates an effect of depth: a stage-wise fidelity curve conflates how hard the
circuit is to execute with how hard the target is to reach. The stage-wise
decomposition into implementation error and incompleteness, the support leakage
and the stage marginals are reported in
Supplement~\supref{sup:stages}, where their limits are set out at length.

What the truncated stages do give is one observable with a known ideal value.
At $L0$ the circuit applies a single rotation to $q_0$; qubits~$q_1$ and~$q_2$
receive no gate at all and are ideally measured as $0$ with probability one, so
any measured $P(q_i{=}1)$ is leakage. At $L01$ the same holds for~$q_2$.

This observable is not, however, independent of the target, and we do not read
it as though it were. The one rotation $L0$ applies carries the target's root
angle, which takes \genRootAngleN{} distinct values across the suite, from
$\genRootAngleLo\dg$ to $\genRootAngleHi\dg$; the residual-coupling and
cross-excitation mechanisms this paper entertains elsewhere are exactly the
ones that would carry that difference to the idle qubits. Variation in the
observable across the full suite is therefore consistent with a changing
instrument state without isolating one.

The suite does, however, contain the comparison that isolates it.
$\genLzeroSubsetN$ of the seven targets --- \genLzeroSubsetTargets{} --- have a
root angle of exactly $\genLzeroSubsetRoot\dg$, so their $L0$ circuits are the
same single gate; this follows from the deposited laws, and the instrument's
own gate lists confirm it for the two of them the export covers. Restricted to
those four sessions the leakage still differs, from $\genLzeroSubsetLo$ to
$\genLzeroSubsetHi$, a factor of $\genLzeroSubsetRatio$, at
$p=\genLzeroSubsetKW$. With the executed circuit held fixed, the instrument is
not in the same state from session to session.

\begin{table}[htbp]
\caption{Population measured on qubits that receive no gate. At $L0$ the circuit
touches only $q_0$, so the ideal value for $q_1$ and $q_2$ is zero; at $L01$ it
touches $q_0$ and $q_1$, so the ideal value for $q_2$ is zero. Entries are means
over the $\genNreps$ runs of each session with sample standard deviations. The
ideal value is zero for every target, but the executed circuit is not the same
for all of them, so variation down a column is not by itself instrument
variation. The attribution in the text is made at $L0$ only, and only for the
\genLzeroSubsetN{} targets whose shared root angle makes the $L0$ circuit a
single identical gate; their $L01$ circuits still differ, so the $L01$ column
carries no such comparison.}
\label{tab:spam}
\centering\small
\begin{tabular}{lccc}
\toprule
Session & \multicolumn{2}{c}{$L0$: $P(q_i{=}1)$, no gate applied} & $L01$: $P(q_2{=}1)$ \\
\cmidrule(lr){2-3}\cmidrule(lr){4-4}
 & $q_1$ & $q_2$ & $q_2$ \\
\midrule
D0 & $0.069 \pm 0.015$ & $0.053 \pm 0.022$ & $0.137 \pm 0.025$ \\
D1 & $0.069 \pm 0.021$ & $0.046 \pm 0.019$ & $0.136 \pm 0.023$ \\
D2 & $0.076 \pm 0.017$ & $0.069 \pm 0.020$ & $0.159 \pm 0.013$ \\
D3 & $0.056 \pm 0.011$ & $0.039 \pm 0.014$ & $0.098 \pm 0.028$ \\
D4 & $0.084 \pm 0.021$ & $0.062 \pm 0.021$ & $0.110 \pm 0.023$ \\
D5 & $0.081 \pm 0.013$ & $0.069 \pm 0.014$ & $0.163 \pm 0.015$ \\
D6 & $0.076 \pm 0.020$ & $0.068 \pm 0.021$ & $0.095 \pm 0.010$ \\
\midrule
Pooled & $0.073$ & $0.058$ & $0.128$ \\
\bottomrule
\end{tabular}

\end{table}

Pooled over the suite this spurious population is $\genSpamQone$ on $q_1$ and
$\genSpamQtwo$ on $q_2$ at $L0$, and $\genSpamQtwoLone$ on $q_2$ at $L01$. The
population falling outside the ideal support runs from $\genLeakLzeroLo$ to
$\genLeakLzeroHi$ at $L0$ and from $\genLeakLoneLo$ to $\genLeakLoneHi$ at
$L01$. We call it an effective state-preparation-and-measurement error, the
adjective doing real work: these data do not separate imperfect preparation,
residual inter-qubit coupling and readout assignment error.

This quantity, whose ideal value is zero for every target, is \emph{not}
constant across the seven sessions. A
Kruskal--Wallis test across sessions gives $p=\genSpamKWqone$ for $q_1$ and
$p=\genSpamKWqtwo$ for $q_2$ at $L0$, and $p=\genSpamKWLone$ at $L01$; the
two-qubit mean at $L0$ runs from $\genSpamRangeLo$ to $\genSpamRangeHi$, a
factor of $\genSpamRangeRatio$. Across all seven this mixes instrument
variation with a possible dependence on the root angle; the
$\genLzeroSubsetRoot\dg$ subset above separates the two, and the instrument
variation survives on its own.

It does not, however, order the targets. The per-session value correlates with
FULL-stage fidelity at $\rho=\genSpamFidRho$ ($p=\genSpamFidP$) and with
execution order at $\rho=\genSpamOrderRho$ ($p=\genSpamOrderP$). With seven sessions
these tests are weak, and the conclusion is correspondingly weak: a drifting
readout floor, the most obvious session-level mechanism, does not track the
fidelity differences among targets. That is not a demonstration that the
session confound is absent.

\subsection{Provenance and the instrument's own record}
\label{subsec:provenance}

The SpinQuasar job database of the CEU unit was exported and is deposited with
this work. It is a second record of the same executions, written by the
instrument rather than by our orchestration script, and neither derives from the
other. For each job it carries the compiled gate list actually executed, the
measured populations and an $8\times8$ complex matrix
(Section~\ref{sec:density-record}).

For the October campaign the cross-check is complete. All $100$ runs are present;
the measured distributions agree with our own record to $5.6\times10^{-17}$, the
largest deviation over the campaign; and the instrument's gate census
independently confirms $21$ gates for every Control run and $18$ for every
Reduced run, which is a check on our compiler and not merely on our logging. The
export additionally holds seven jobs carrying campaign run names from an aborted
first attempt, whose resume logic keyed on run name alone; that attempt lost its
connection after seven jobs and recorded nothing, the campaign was reacquired
from scratch, and each deposited run matches the later of the two device jobs to
floating-point identity. The acquisition script now refuses to continue when a
log holds runs from a different backend.

For the benchmark campaign the cross-check is \emph{partial}, and this is a
reproducibility gap rather than a formality. The export contains $498$ of the
$\genSessNtotal$ runs. It has no entry at all for $31$ March, the day on which
the D3 and D2 sessions ran, so those $200$ runs have no instrument-side record;
two further runs are absent from an otherwise complete $1$ April export. This
was established rather than assumed: each target commands a distinct set of
Gray-code sub-rotations, recoverable from the deposited gate lists, and
fingerprinting all $986$ circuits in the export recovers D1, D4, D5 and D6 while
returning nothing for D2 or D3 under any job name. Three further properties of the
export matter to anyone using it: the device clock runs two hours ahead of our
UTC timestamps, uniformly and to the second; there is no shot count anywhere in
it; and its own \texttt{Fidelity} column is empty for every row.

Every numerical table in this paper, and every figure on which a claim rests,
is generated from the deposited run-level data by a single script.

\section{Characterisation across the benchmark suite}
\label{sec:suite}

The first campaign asks whether the two Gray-code realisations perform alike
across a suite of targets, and bounds the answer.

\subsection{The campaign}
\label{subsec:suite-design}

Seven targets, three stages and two ladder realisations were executed over
$700$ runs between 31~March and 2~April 2026, $25$ runs in each
(target, stage, ladder) cell. The targets span a deliberate range of entropy,
contrast and dyadic-tree structure, and are defined with their verification
digests in Section~\ref{sec:target}. The truncated stages $L0$ and $L01$ are
auxiliary controls in the sense of Section~\ref{subsec:stages}; everything in
this section refers to the FULL stage, where the circuit implements the complete
preparation and the ideal output is the target itself. The stage-wise material
is reported in Supplement~\supref{sup:stages}.

Pooled over both ladders, mean FULL-stage fidelity ranges from $0.871$ for D3 to
$0.964$ for D6. We report means, dispersions and intervals throughout; the best
individual run of a cell is given in the tables for completeness but carries no
weight in any comparison, since a maximum over $25$ draws is a statement about
the tail and not about the preparation.

\subsection{The two realisations do not perform alike}
\label{subsec:suite-ab}

\begin{table}[htbp]
\centering\small
\caption{FULL-stage performance for every target and both ladders, with the A/B
contrast under three multiplicity criteria. Raw $p$-values are two-sided
Mann--Whitney \cite{MannWhitney1947} over $25$ runs per cell. Holm controls the family-wise error rate
under any dependence; Benjamini--Hochberg controls the false discovery rate
under independence or positive regression dependence; Benjamini--Yekutieli
controls it under arbitrary dependence, at a cost of
$c(7)=\genBYconst$.}
\label{tab:full_ab}
\begin{tabular}{llcccccc}
\toprule
Target & Ladder & Fidelity & $\TV$ & $p_{\mathrm{raw}}$ & $p_{\mathrm{Holm}}$ & $p_{\mathrm{BH}}$ & $p_{\mathrm{BY}}$ \\
\midrule
D0 & A & $0.954 \pm 0.014$ & $0.163 \pm 0.030$ & \multirow{2}{*}{$0.0002$***} & \multirow{2}{*}{$0.0014$} & \multirow{2}{*}{$0.0008$}& \multirow{2}{*}{$0.0021$} \\
   & B & $0.970 \pm 0.014$ & $0.139 \pm 0.034$ & & & & \\
\addlinespace
D1 & A & $0.908 \pm 0.019$ & $0.268 \pm 0.031$ & \multirow{2}{*}{$0.801$\,ns} & \multirow{2}{*}{$1.000$} & \multirow{2}{*}{$0.846$}& \multirow{2}{*}{$1.000$} \\
   & B & $0.909 \pm 0.017$ & $0.267 \pm 0.023$ & & & & \\
\addlinespace
D2 & A & $0.950 \pm 0.018$ & $0.140 \pm 0.026$ & \multirow{2}{*}{$0.112$\,ns} & \multirow{2}{*}{$0.335$} & \multirow{2}{*}{$0.156$}& \multirow{2}{*}{$0.405$} \\
   & B & $0.957 \pm 0.022$ & $0.127 \pm 0.039$ & & & & \\
\addlinespace
D3 & A & $0.859 \pm 0.039$ & $0.335 \pm 0.056$ & \multirow{2}{*}{$0.014$*} & \multirow{2}{*}{$0.058$} & \multirow{2}{*}{$0.025$}& \multirow{2}{*}{$0.066$} \\
   & B & $0.883 \pm 0.018$ & $0.303 \pm 0.028$ & & & & \\
\addlinespace
D4 & A & $0.882 \pm 0.045$ & $0.290 \pm 0.056$ & \multirow{2}{*}{$0.846$\,ns} & \multirow{2}{*}{$1.000$} & \multirow{2}{*}{$0.846$}& \multirow{2}{*}{$1.000$} \\
   & B & $0.882 \pm 0.029$ & $0.293 \pm 0.041$ & & & & \\
\addlinespace
D5 & A & $0.951 \pm 0.012$ & $0.115 \pm 0.031$ & \multirow{2}{*}{$0.0019$**} & \multirow{2}{*}{$0.0095$} & \multirow{2}{*}{$0.0044$}& \multirow{2}{*}{$0.012$} \\
   & B & $0.934 \pm 0.022$ & $0.144 \pm 0.039$ & & & & \\
\addlinespace
D6 & A & $0.950 \pm 0.016$ & $0.198 \pm 0.034$ & \multirow{2}{*}{$<10^{-4}$***} & \multirow{2}{*}{$<10^{-4}$} & \multirow{2}{*}{$<10^{-4}$}& \multirow{2}{*}{$<10^{-4}$} \\
   & B & $0.979 \pm 0.007$ & $0.108 \pm 0.020$ & & & & \\
\addlinespace
\bottomrule
\end{tabular}

\end{table}

Seven contrasts are tested, so a multiplicity criterion is required. Under
Holm \cite{Holm1979} the contrast is significant for \genNsigHolm{} targets
(\genSigHolmList); under Benjamini--Hochberg \cite{BenjaminiHochberg1995} for
\genNsigBH{} (\genSigBHList). The single target that changes status is
$\genMultFlipTarget$, whose raw $p$ of $\genMultFlipRaw$ Holm inflates to
$\genMultFlipHolm$ and Benjamini--Hochberg leaves at $\genMultFlipBH$.

The choice between them is not free here. Benjamini--Hochberg requires
independence or positive regression dependence, and the seven targets were
acquired in seven contiguous sessions on one instrument, so drift couples the
contrasts. Benjamini--Yekutieli \cite{BenjaminiYekutieli2001}, which assumes
nothing about the dependence structure, returns \genNsigBY{} targets
(\genSigBYList) --- the same set as Holm. \textbf{Both procedures that make no
independence assumption select the same three targets; the count of four arises
only under the procedure whose validity rests on an assumption this design does
not guarantee.} We take Holm as the primary criterion and report the others
alongside it; Supplement~\supref{sup:multiplicity} gives the procedures, the
ordered table and the dependence argument in full.

\paragraph{The sign is not fixed.} Among the \genNsigHolm{} targets Holm
selects, D0 and D6 favour ladder~B while D5 favours ladder~A by $0.018$; D3,
which only Benjamini--Hochberg selects, also favours ladder~B.
A difference of fixed sign would admit a simple reading --- that one of the two
Gray-code orderings is the better compilation --- and that reading is excluded.
Whatever produces the contrast interacts with something that varies across the
suite.

\subsection{The uniform target isolates scheduling from angle content}
\label{subsec:suite-d0}

D0 is the most informative single row of Table~\ref{tab:full_ab}, and it is
worth separating from the rest.

For the uniform distribution all four level-2 conditional rotations are equal,
so the Walsh--Hadamard inversion that produces the ladder coefficients leaves a
single non-zero sub-rotation. Both realisations then command the identical angle
sequence $(90\dg{},0,0,0)$: not merely the same unitary, which is true for every
target, but the same numbers sent to the same gates. The compiled circuits
differ \emph{only} in which qubit controls each CNOT.

Yet the contrast on D0 is the second strongest of the suite, significant under
all three criteria. Whatever separates the two realisations therefore has a
component that does not depend on the rotation angles at all, and that must be
carried by the order in which the operations are scheduled. The asymmetric
targets, where the two ladders command genuinely different sub-rotation
sequences, may add an angle-dependent component on top; D0 shows that a
scheduling component exists on its own.

\subsection{Tree asymmetry: an exploratory association}
\label{subsec:suite-asymmetry}

Pooled FULL-stage fidelity is associated with the asymmetry of the underlying
dyadic tree rather than with Shannon entropy or contrast. The rank correlation
with the level-1 deviation $\Delta_1$ is $\rho=-0.64$ with $p=0.12$, against
$\rho=+0.39$ for entropy and $\rho=-0.11$ for contrast. A matched comparison
helps: among four targets lying within $0.014$ of each other in distance from
the uniform law, fidelity still spans $0.882$ to $0.964$ and remains ordered by
tree asymmetry.

This is exploratory and is not a predictor. There are seven independent
targets, however many runs each contains, and $p=0.12$ does not establish an
association; the two asymmetry descriptors are collinear at
$\rho=0.93$ and are two parametrisations of one feature rather than two pieces
of evidence; and the matched comparison does not break the session confound,
because its four targets come from four different sessions. The defensible
statement is that this suite shows an exploratory association between tree
asymmetry and performance, requiring validation in interleaved acquisition.
Descriptor definitions and the full analysis are in
Supplement~\supref{sup:asymmetry}.

\subsection{Correlations the single-qubit marginals cannot see}
\label{subsec:correlators}

The comparisons above read the measured population vector through a scalar
distance to the target. That is the quantity the application cares about, and it
is a lenient one: on~D1 a classical fidelity of $\genLenientFid$ coexists with a
total-variation distance of $\genLenientTV$, so a quarter of the probability
mass sits in the wrong place behind a figure of merit above~$0.9$. The same
counts support other readings, one of which costs nothing to obtain and
constrains what the errors can be.

For a measured law $q$ on $\{0,1\}^3$ the quantities
\begin{equation}
\langle Z_iZ_j\rangle_q=\sum_{x}q(x)\,(-1)^{x_i+x_j},
\qquad
\langle Z_0Z_1Z_2\rangle_q=\sum_{x}q(x)\,(-1)^{x_0+x_1+x_2},
\label{eq:correlators}
\end{equation}
are linear functionals of the same population vector that
Section~\ref{sec:design} already reports. They are not a second experiment: no
basis change is performed, no reconstruction is involved, and no assumption is
made beyond those under which the distributions themselves are read. Being
diagonal in the computational basis, they inherit unchanged whatever confounding
of preparation with readout the distributions carry. What they add is not
independence from readout but sensitivity to structure that the single-qubit
marginals are blind to.

That blindness is not incidental for this benchmark suite:
$\genNbalancedTargets$ of the seven targets --- D0, D1, D3 and D4 --- have all
three single-qubit marginals exactly equal to $1/2$. For those targets the
marginals carry no information whatever: any state with balanced marginals
reproduces them exactly, and everything that distinguishes the target from the
uniform law is held in the correlations. A report of marginal balance for such a
target, of the kind Section~\ref{subsec:int-marginals} gives for~D1, establishes
that nothing has gone wrong with the single-qubit populations and nothing more.

\begin{table}[htbp]
\caption{Two- and three-body correlators of
Equation~\eqref{eq:correlators} for the seven targets of the benchmark suite,
target against measured. Measured values are the correlators of the mean
measured law over the $50$ FULL runs of each target, both ladders pooled;
per-run distributions and the ladder-resolved breakdown are in
Supplement~\supref{sup:correlators-suite}.}
\label{tab:correlators_suite}
\centering\small
\begin{tabular}{llcccc}
\toprule
Target & & $\langle Z_0Z_1\rangle$ & $\langle Z_0Z_2\rangle$ & $\langle Z_1Z_2\rangle$ & $\langle Z_0Z_1Z_2\rangle$ \\
\midrule
D0 & target & $+0.000$ & $+0.000$ & $+0.000$ & $+0.000$ \\
   & measured & $+0.217$ & $+0.002$ & $+0.060$ & $-0.067$ \\
\addlinespace
D1 & target & $-0.400$ & $-0.200$ & $+0.000$ & $+0.000$ \\
   & measured & $+0.131$ & $-0.146$ & $-0.063$ & $-0.112$ \\
\addlinespace
D2 & target & $+0.280$ & $+0.100$ & $+0.060$ & $+0.040$ \\
   & measured & $+0.358$ & $+0.117$ & $+0.020$ & $+0.014$ \\
\addlinespace
D3 & target & $-0.800$ & $-0.720$ & $+0.680$ & $+0.000$ \\
   & measured & $-0.327$ & $-0.398$ & $+0.200$ & $+0.069$ \\
\addlinespace
D4 & target & $-0.720$ & $-0.440$ & $+0.320$ & $+0.000$ \\
   & measured & $-0.160$ & $-0.289$ & $+0.035$ & $-0.041$ \\
\addlinespace
D5 & target & $+0.512$ & $-0.329$ & $+0.003$ & $+0.141$ \\
   & measured & $+0.279$ & $-0.290$ & $-0.039$ & $+0.169$ \\
\addlinespace
D6 & target & $+0.126$ & $-0.353$ & $-0.400$ & $+0.348$ \\
   & measured & $+0.148$ & $-0.303$ & $-0.234$ & $+0.304$ \\
\addlinespace
\bottomrule
\end{tabular}

\end{table}

Table~\ref{tab:correlators_suite} gives the comparison. Three features of it
bear on the argument of this paper.

First, marginal accuracy and correlation accuracy are not the same measurement.
D4 has the most accurately reproduced marginals in the suite --- all three
within $\genMargDevDfour$ of their target values --- and reproduces its dominant
two-body correlator at $\genCorrRatioDfour\%$ of the target value
($\genCorrTgtDfour$ against $\genCorrObsDfour$). D3 reproduces its marginals to
$\genMargDevDthree$ and its dominant correlator at $\genCorrRatioDthree\%$. On
these targets a report of marginal balance and a report of the measured
correlations would support opposite impressions of how well the state was
prepared.

Second, structure appears where the target has none. D0 is the uniform law: all
four correlators of Equation~\eqref{eq:correlators} vanish identically, and D0
attains a classical fidelity of $\genCorrDzeroF$, among the highest in the
suite. Its measured $\langle Z_0Z_1\rangle$ is $\genCorrDzeroZZ$, while the
other two pair correlators are $+0.002$ and $+0.060$. A correlation of that size
appears on a target that has none, it is confined to one qubit pair, and no
single-qubit quantity registers it.

Third, these two observations together exclude the simplest account of the
measured laws. If the device acted as a global depolarising channel --- if the
measured law were $\lambda p^{\mathrm{ideal}}+(1-\lambda)u$ with $u$ uniform ---
then every correlator and every marginal deviation would be the target value
scaled by one common factor~$\lambda$, and an observable whose target value is
zero would be measured at zero. Fitting a single $\lambda$ across all $49$
target--observable pairs of the suite (four correlators and three marginals per
target) gives $\lambda=\genCorrLambda$ with $R^2=\genCorrRtwo$ and a largest
residual of $\genCorrResidMax$. Allowing a separate $\lambda$ per target does not
rescue the description: the fitted values range from $\genCorrLamMin$ on~D1,
where the dominant correlation is measured with the wrong sign, to
$\genCorrLamMax$ on~D2, where the measured correlations are \emph{larger} in
magnitude than the target's. And of the $\genCorrZeroN$ observables whose target
value is exactly zero, the measured values reach $\genCorrZeroMax$ in magnitude,
with a root-mean-square of $\genCorrZeroRms$.

The conclusion is negative. A description in
which the preparation is correct and the output is subsequently mixed towards
the uniform law is not consistent with these data: the errors are not a
contraction of the target's structure but a redistribution of it, and their
pattern depends on which qubit pair is involved. This is a constraint on
candidate mechanisms, taken up in Section~\ref{sec:discussion}; it is not an
identification of one.

\subsection{What this design cannot separate}
\label{subsec:suite-limits}

Each target was acquired in its own contiguous session, so target is aliased
with session: nothing in this campaign can distinguish a property of a
distribution from a property of the hours in which it was measured. The
execution order was D3, D2, D5, D6, D4, D1, D0, and pooled fidelity does not
degrade monotonically along it ($\rho=+0.32$, $p=0.48$), which excludes a simple
monotone deterioration but not non-monotone changes of the instrument.

The A/B contrast is the one comparison this design makes within a session, which
is why it is the quantity this section reports. Even so, the ladders were
acquired in alternating blocks of five rather than interleaved, so within each
session ladder~B sits a mean of five slots later than ladder~A; the
consequences are quantified in Section~\ref{subsec:long-caveats}.

These limitations are what the campaign of Section~\ref{sec:intervention} was
designed to remove: it fixes the target, interleaves the realisations run by
run, and introduces an intervention whose effect on each can be compared within
repetitions.

\section{Intervention within a single unit}
\label{sec:intervention}

Section~\ref{sec:suite} can say no more than that the two realisations differ,
because its design varies only the target: each distribution has its own
session, so target and session are aliased.

This campaign was designed to go further. It holds the target
fixed and introduces a controlled intervention on the circuit, applied to both
realisations, so that the question becomes whether the two compilations respond
to the same modification in the same way. They do not, and the size of the
difference between their responses is the principal quantitative result of this
paper.

\subsection{Design}
\label{subsec:int-design}

The campaign was executed on the CEU unit, on the FULL stage of the D1
benchmark, under the stock pulse calibration of the device. Two circuit
configurations were compared:

\begin{itemize}[leftmargin=*,itemsep=3pt]
\item \textbf{Control.} The compiled circuit as published here: $21$ gates
  ($11\,\Ry$, $8\,\mathrm{CNOT}$, $2\,X$), retaining the $\Ry(0)$ placeholder at
  position $a_2$ of the level-2 ladder and the identity-safe tail on $q_0$.
\item \textbf{Reduced.} The same logical circuit with those identity-equivalent
  gates removed --- the zero-angle rotation and the two-gate tail --- giving
  $18$ gates ($8\,\Ry$, $8\,\mathrm{CNOT}$, $2\,X$). The pulse calibration is
  unchanged, and the unitary is unchanged.
\end{itemize}

The intervention is deliberately minimal. It removes three gates whose combined
action is the identity, so the implemented transformation is exactly preserved
and the only thing that changes is the sequence the instrument executes. It is
also identical for the two ladders: the zero-angle rotation sits at position
$a_2$ in both, since for D1 the level-2 sub-rotation sequences are
$(90\dg{},\,5.63\dg{},\,0,\,13.84\dg{})$ for ladder~A and
$(90\dg{},\,13.84\dg{},\,0,\,5.63\dg{})$ for ladder~B, and the tail is the same
pair of gates in each. Whatever difference the two realisations show in their
response cannot therefore be attributed to a difference in what was removed.

Each configuration was executed with both ladders, $25$ runs per cell at $2048$
nominal shots: $100$ hardware runs in a single session of $3.72$~h. Three features of
the acquisition carry the comparison.

\paragraph{Interleaving.} The four cells were interleaved rather than blocked,
with the order within each repetition reversed on alternate repetitions. The
cells occupy mean positions $50.4$, $50.5$, $50.5$ and $50.6$ of $100$, so no
cell systematically occupies an early or a late slot. This is the confound that
the campaign of Section~\ref{sec:suite} could not avoid.

\paragraph{Pairing.} Because the cells occupy adjacent slots within each
repetition, every contrast can be formed \emph{within} a repetition and tested
as a paired comparison. Drift slower than the $126$~s acquisition cadence is strongly suppressed,
though not cancelled: the cells are adjacent, not simultaneous. We report
paired tests throughout; the unpaired versions agree, and
Supplement~\supref{sup:intervention-unpaired} gives both, together with the
serial-dependence treatment the pairing does not cover.

\paragraph{Pre-acquisition verification.} Each of the four circuit variants was
simulated against the analytic target before any hardware job was submitted, and
the acquisition script refuses to proceed unless all four reproduce it; the
measured agreement was $\TV \le 1.4\times10^{-10}$ in every case. The bit-order
convention of the unit was measured fifteen minutes before the session and is
recorded in the instrument log.

The session was interrupted once by a loss of connection and resumed after
$17.4$~min. The acquisition before and after the break does not differ
($0.883 \pm 0.054$ against $0.896 \pm 0.042$, Welch \cite{Welch1947} $p=0.18$), and fidelity shows
no monotone trend with run order ($\rho=+0.13$, $p=0.20$).

\subsection{The two realisations respond differently to the same intervention}
\label{subsec:int-results}

\begin{table}[htbp]
\caption{The four cells of the October factorial. Entries are means over
$\genNreps$ runs with sample standard deviations; the last column is the best
single run in the cell.}
\label{tab:october_cells}
\centering\small
\begin{tabular}{llcccc}
\toprule
Configuration & Ladder & Fidelity & $\TV$ & $\ell^2$ & Fidelity max \\
\midrule
Control  & A & $0.886 \pm 0.054$ & $0.283 \pm 0.082$ & $0.227 \pm 0.063$ & $0.967$ \\
         & B & $0.847 \pm 0.035$ & $0.334 \pm 0.055$ & $0.271 \pm 0.029$ & $0.916$ \\
\addlinespace
Reduced  & A & $0.900 \pm 0.045$ & $0.244 \pm 0.058$ & $0.210 \pm 0.047$ & $0.974$ \\
         & B & $0.923 \pm 0.025$ & $0.234 \pm 0.040$ & $0.195 \pm 0.032$ & $0.965$ \\
\addlinespace
\bottomrule
\end{tabular}

\end{table}

Under stock calibration the two ladders differ by $0.040$ in mean fidelity, with
ladder~A ahead. Removing the three identity-equivalent gates raises the pooled
mean from $0.867$ to $0.912$. Neither of those statements is the result; the
result is that the improvement is not shared.

\begin{table}[htbp]
\caption{Paired contrasts over the $\genNreps$ repetitions. Each row compares
two cells acquired under the same conditions within each repetition, so the
intervals do not rest on treating runs within a cell as independent draws; the
signed-rank column is Wilcoxon's \cite{Wilcoxon1945}, and the
unpaired counterparts are in Supplement~\supref{sup:intervention-unpaired}.}
\label{tab:october_contrasts}
\centering\small
\begin{tabular}{lcccc}
\toprule
Paired contrast & Estimate & 95\% CI & $p$ (paired $t$) & $p$ (Wilcoxon) \\
\midrule
Ladder A $-$ B, Control & $+0.0395$ & $[+0.0165,\,+0.0626]$ & $0.0017$ & $0.0023$ \\
Ladder A $-$ B, Reduced & $-0.0229$ & $[-0.0431,\,-0.0027]$ & $0.028$ & $0.039$ \\
Reduction effect, ladder~A & $+0.0140$ & $[-0.0131,\,+0.0411]$ & $0.296$ & $0.210$ \\
Reduction effect, ladder~B & $+0.0765$ & $[+0.0573,\,+0.0957]$ & $<10^{-4}$ & $<10^{-4}$ \\
Ladder $\times$ reduction interaction & $+0.0625$ & $[+0.0306,\,+0.0944]$ & $0.0005$ & $0.0001$ \\
\bottomrule
\end{tabular}

\end{table}

Ladder~B gains $0.077$ from the removal, with a 95\% confidence interval
excluding zero by a wide margin. Ladder~A gains $0.014$, with an interval that
includes zero. The difference between those two responses --- the
ladder~$\times$~intervention interaction --- is $\genInterEst$ with 95\% CI
$[\genInterLo,\genInterHi]$, paired $p=\genInterP$ over \genNreps{}
repetitions. The same three gates, removed from two circuits that implement the
same unitary, produce effects that differ by a factor of five.

The intervention also does not preserve the ordering of the two realisations.
In the Control configuration
ladder~A leads; in the Reduced configuration ladder~B leads by $0.023$ (paired
$p=0.028$), and the two mean laws move towards each other, $\TV(A,B)$ falling
from $0.093$ to $0.060$ and $\Fid(A,B)$ rising from $0.983$ to $0.994$. A
modification that preserves the unitary exactly therefore reverses which
compilation performs better.

One qualification belongs with that statement. The interaction is the contrast
this campaign was designed to estimate and the only one specified in advance;
the other four rows of Table~\ref{tab:october_contrasts} are secondary readings
of the same design. The reversal rests on the Reduced contrast, and that
contrast does not survive a Holm correction applied across all five rows,
whereas the interaction and the Control contrast do. The change of sign in the
estimates is what we report; its significance as a separate finding is not
established here.

\subsection{A scheduling asymmetry in the executed circuits}
\label{subsec:int-schedule}

The instrument records the gate list it actually executed for every job, and
those records make the asymmetry concrete rather than inferred. Each of the four
cells executed a single, identical sequence across all of its runs. In the
Control configuration the identity-safe tail is \emph{placed differently} in the
two ladders: for ladder~A the two tail rotations are appended at the end of the
circuit, at positions $19$ and $20$ of $21$; for ladder~B they occupy positions
$18$ and $20$, separated by the final CNOT of the level-2 ladder.

Removing the tail therefore does not do the same thing to the two circuits. In
ladder~A it truncates two gates from the end of a sequence that is otherwise
untouched. In ladder~B it also removes a gate from \emph{between} two
two-qubit operations, changing what the final CNOT is adjacent to. This is
observable in the instrument's own record, it is consistent across all $50$
Control runs, and it is the kind of difference the logical description of the
circuit does not represent at all.

We offer it as the most economical account of the interaction, not as an
identification. Establishing it would require separating the two removed
elements, which the present design does not do: Section~\ref{sec:discussion}
sets out the comparison that would.

\subsection{What the intervention does not fix}
\label{subsec:int-marginals}

\begin{center}\small
\begin{tabular}{llccc}
\toprule
Configuration & Ladder & $P(q_0{=}1)$ & $P(q_1{=}1)$ & $P(q_2{=}1)$ \\
\midrule
Control  & A & $0.498$ & $0.481$ & $0.515$ \\
         & B & $0.489$ & $0.394$ & $0.496$ \\
\addlinespace
Reduced  & A & $0.529$ & $0.488$ & $0.438$ \\
         & B & $0.504$ & $0.504$ & $0.497$ \\
\addlinespace
\midrule
Mar--Apr 2026 & A & $0.524$ & $0.491$ & $0.450$ \\
\bottomrule
\end{tabular}

\end{center}

Under stock calibration the single-qubit marginals of ladder~A are close to
balanced, while ladder~B carries a pronounced deficit on $q_1$,
$P(q_1{=}1)=0.394$ against an ideal $0.5$. The intervention restores that
marginal to $0.504$, which is consistent with the deficit being carried by the
compiled schedule rather than by the logical circuit.

It does not restore balance across the board. In the Reduced configuration
ladder~A shows $P(q_2{=}1)=0.438$, a deviation of $0.062$ that is larger than
any single deviation of Control ladder~A. Improving the joint distribution and
improving the single-qubit marginals are not the same thing here, and a
configuration can do the first while worsening the second.

\subsection{The intervention seen in a correlator}
\label{subsec:int-correlator}

Section~\ref{subsec:correlators} reads a measured law through the two- and
three-body correlators of Equation~\eqref{eq:correlators}. Applied to this
campaign, with the same $100$ runs and the same pairing by repetition, it gives
a second view of what the intervention did.

\begin{table}[htbp]
\caption{The October intervention read through $\langle Z_0Z_1\rangle$, the
dominant correlator of~D1 (target value $\genCorrTargetZZ$). Entries are means
over the $\genNreps$ repetitions with sample standard deviations; the last
column is the paired Reduced\,$-$\,Control difference with its $95\%$ interval
and the paired $t$-test $p$-value.}
\label{tab:correlator_intervention}
\centering\small
\begin{tabular}{lccc}
\toprule
Ladder & Control & Reduced & Reduced $-$ Control \\
\midrule
A & $+0.160 \pm 0.169$ & $+0.046 \pm 0.135$ & $-0.113$ $[-0.205,-0.022]$, $0.017$ \\
B & $+0.217 \pm 0.175$ & $+0.010 \pm 0.105$ & $-0.207$ $[-0.278,-0.135]$, $3.8\times10^{-6}$ \\
\bottomrule
\end{tabular}

\end{table}

Removing the three gates whose joint action is the identity moves this
correlator towards its target value in both realisations, and by different
amounts: by
$\genCorrRedAEst$ $[\genCorrRedALo,\genCorrRedAHi]$ for ladder~A
($p=\genCorrRedAP$) and by $\genCorrRedBEst$
$[\genCorrRedBLo,\genCorrRedBHi]$ for ladder~B ($p=\genCorrRedBP$). The
direction of the asymmetry is the one reported for the classical fidelity in
Section~\ref{subsec:int-results}: ladder~B responds to the intervention more
strongly than ladder~A.

This is not independent evidence that the intervention acted differently on the
two realisations: it is computed from the same counts as the fidelity analysis,
and a second functional of one data set is not a second experiment. What it adds
is information about \emph{what} moved. The effect of the intervention is not
distributed evenly over the law: it is concentrated in the correlator that
dominates this target. We do not read that as localising the effect to a
particular pair of qubits. The removed gates act on $q_0$, and the CNOTs of the
level-2 ladder couple $q_0$ and $q_1$ to $q_2$, so $\langle Z_0Z_1\rangle$ is
not the pair those gates touch; it is the component of the law in which D1
differs most from the uniform distribution.

The improvement is real but partial. The residual values are $\genCorrRedARes$
and $\genCorrRedBRes$ against a target of $\genCorrTargetZZ$: both realisations
move towards the target, neither reaches it, and both still carry the dominant
correlation of the target with the wrong sign after the circuits have been
shortened. A gate-count reduction that raises the classical fidelity by a
measurable amount reduces the discrepancy in the respect that distinguishes
this target from the uniform law without correcting its sign.

\subsection{Corroboration and scope}
\label{subsec:int-scope}

The instrument's job database was exported after the session as a second and
independent record of the same executions. All $100$ runs are present in it and
the measured distributions agree with our own record to floating-point identity,
the largest deviation over the campaign being $5.6\times10^{-17}$. The
instrument's gate census independently confirms $21$ gates for every Control run
and $18$ for every Reduced run. This is a traceability check on labelling and
logging; it is documentary corroboration that the two records describe the same
acquisitions, and not a second physical measurement.

Three limits bound what this section establishes. The intervention removes two
kinds of gate at once --- the zero-angle rotation and the tail --- and measures
only their joint effect. The campaign uses one target, so nothing here speaks to
how the interaction varies across the benchmark suite. And the comparison is
made under one compilation of one circuit on one unit;
Section~\ref{sec:compilations} shows that the first of these matters.

A fourth limit is a property of the instrument rather than of the design.
Triangulum reads out an ensemble, so a run carries no sampling scale of its own
(Section~\ref{subsec:acquisition}) and the variation we can measure is the
variation between executions. That is what the paired design uses and what
every interval here is computed from;
Supplement~\supref{sup:intervention-unpaired} sets out what the paired and
unpaired analyses each assume.

\section{The ladder contrast across compilations of one circuit}
\label{sec:compilations}

Section~\ref{sec:intervention} compares two realisations under one circuit
intervention, in one session. This section widens it along the two axes that campaign holds fixed: the
compilation of a circuit whose unitary never changes, and the passage of time
with the compilation held fixed. Both were forced on us by a discrepancy in the
instrument's own record, and one required a new experiment.

\subsection{A discrepancy in the instrument's record}
\label{subsec:long-three}

The D1 benchmark appears in both campaigns, and we initially treated the two
acquisitions as running the same circuit six months apart. The instrument's job
database shows that they did not. For the same target, from the same source, the
compiler emitted two different gate lists.

The first difference is the identity-safe tail of
Section~\ref{subsec:int-design}: the April circuit does not carry it, so it has
$\genCmpAprGates$ gates where the October Control circuit has
$\genCmpCtlGates$. The second is a convention. The level-1 uniformly controlled
rotation is realised as
$\Ry(+\theta/2)\,\mathrm{CNOT}\,\Ry(-\theta/2)\,\mathrm{CNOT}$, and the April
compiler emitted the second rotation not as $-\theta/2$ but as
$720\dg-\theta/2$. On D1 the two commands read $+663.211\dg$ and $+686.789\dg$
where October commands $-56.789\dg$ and $-33.211\dg$. The convention is
systematic: it appears at the same two circuit positions, on both ladders, for
every target present in the instrument export, with commanded angles between
$651\dg$ and $699\dg$. The October campaign never commands a rotation above
$90\dg$.

These are the same unitary. A $720\dg$ rotation of a spin-$\tfrac12$ is the
identity, phase included, so the two commands implement identical gates and the
two circuits are equal as operators, not merely as maps on the computational
basis. Whether they are the same physical operation is a different question. On
a liquid-state NMR processor the commanded angle sets the pulse area, so a
$663\dg$ rotation should be delivered as a pulse roughly an order of magnitude
longer than the $-57\dg$ one it is equivalent to. That is a question about the
device, not about the mathematics, and Section~\ref{subsec:long-angle} settles
it experimentally rather than by argument.

\begin{table}[htbp]
\caption{The ladder contrast on~D1 across every compiled realisation of this
logical circuit now acquired, ordered by what distinguishes them. Entries are
means over $\genNreps$ runs with sample standard deviations; the last column
gives $A-B$ with the Mann--Whitney $p$-value for that acquisition. The first
three rows are the same gate list to within two angle values; the first two are
the same gate list exactly, two days apart.}
\label{tab:compilations}
\centering\small
\begin{tabular}{llllccc}
\toprule
Acquisition & Gates & Undo rotations & Tail & Ladder A & Ladder B & A $-$ B \\
\midrule
5~Oct, Control & 21 & $-\theta$ & present & $0.886 \pm 0.054$ & $0.847 \pm 0.035$ & $+0.0395$ (0.0038) \\
7~Oct, direct & 21 & $-\theta$ & present & $0.921 \pm 0.030$ & $0.906 \pm 0.041$ & $+0.0146$ (0.269) \\
7~Oct, folded & 21 & $-\theta+720\dg$ & present & $0.906 \pm 0.051$ & $0.921 \pm 0.033$ & $-0.0150$ (0.404) \\
1~Apr, FULL & 19 & $-\theta+720\dg$ & absent & $0.908 \pm 0.019$ & $0.909 \pm 0.017$ & $-0.0014$ (0.801) \\
5~Oct, Reduced & 18 & $-\theta$ & absent & $0.900 \pm 0.045$ & $0.923 \pm 0.025$ & $-0.0229$ (0.050) \\
\bottomrule
\end{tabular}

\end{table}

\subsection{The angle representative: a null, measured}
\label{subsec:long-angle}

The convention can be tested directly. The two arms differ in the
representation convention applied to two angles, and in nothing else that is
commanded: the level-1 inverse rotations, whose values differ from each other,
are commanded as $-\theta$ in one arm and as $720\dg-\theta$ in the other. Same gate list,
same gate count, same gate types, same controls, same every other angle. The
execution times we cannot vouch for, for the reason given below. We acquired it as a $2\times2$ factorial, convention
$\times$ ladder, $\genAngN$ repetitions per cell, in one session, interleaved
run by run and paired by repetition --- the design of
Section~\ref{sec:intervention} with the intervention replaced.

\begin{table}[htbp]
\caption{The angle-representative factorial on~D1, one session, interleaved and
paired. The last column is the commanded value of the two level-1 inverse
rotations; every other gate and angle is common to all four cells, verified
against the gate lists the instrument recorded.}
\label{tab:angle_cells}
\centering\small
\begin{tabular}{llccc}
\toprule
Convention & Ladder & Fidelity & $\TV$ & Undo rotations \\
\midrule
Direct, $-\theta$        & A & $0.921 \pm 0.030$ & $0.222 \pm 0.046$ & $-56.7891\dg$ \\
                         & B & $0.906 \pm 0.041$ & $0.247 \pm 0.063$ & $-56.7891\dg$ \\
\addlinespace
Folded, $-\theta+720\dg$ & A & $0.906 \pm 0.051$ & $0.243 \pm 0.072$ & $663.2109\dg$ \\
                         & B & $0.921 \pm 0.033$ & $0.219 \pm 0.053$ & $663.2109\dg$ \\
\addlinespace
\bottomrule
\end{tabular}

\end{table}

\begin{table}[htbp]
\caption{Paired contrasts over the $\genAngN$ repetitions of the
angle-representative factorial.}
\label{tab:angle_contrasts}
\centering\small
\begin{tabular}{lcccc}
\toprule
Paired contrast & Estimate & 95\% CI & $p$ (paired $t$) & $p$ (Wilcoxon) \\
\midrule
Folded $-$ direct, ladder~A & $-0.0154$ & $[-0.0410,\,+0.0102]$ & $0.227$ & $0.442$ \\
Folded $-$ direct, ladder~B & $+0.0142$ & $[-0.0017,\,+0.0302]$ & $0.078$ & $0.134$ \\
Ladder A $-$ B, direct & $+0.0146$ & $[-0.0058,\,+0.0351]$ & $0.153$ & $0.113$ \\
Ladder A $-$ B, folded & $-0.0150$ & $[-0.0414,\,+0.0114]$ & $0.253$ & $0.442$ \\
Convention $\times$ ladder interaction & $-0.0296$ & $[-0.0606,\,+0.0014]$ & $0.061$ & $0.090$ \\
\bottomrule
\end{tabular}

\end{table}

Two checks come before the result. The instrument's record confirms that the
two arms differ in exactly two gate positions and in nothing else, and that the
circuit the device accepted carries $\genAngCmdFolded\dg$ in the folded arm
rather than a wrapped equivalent. The export also reproduces our own run
records to $6.1\times10^{-16}$.

What that record establishes is the command, not the pulse. We have no
control-layer data --- no duration, amplitude, phase or waveform, and no
manufacturer specification of how a commanded angle is realised --- so that the
two arms were driven differently follows only if the delivered pulse area
tracks the commanded angle, which is an assumption and not an observation. The
same gap qualifies the claim that the schedule is held fixed: if the folded
arm's pulse is longer, the gate list is identical but the execution times are
not. The arms are therefore known to differ in what was commanded, and
plausibly in what was delivered, with the timing consequence undetermined.

The result is a null. Pooled over ladders, the folded arm differs from the
direct arm by $-0.0006$ with a $95\%$ interval of $[-0.0157,+0.0146]$ and
$p=0.94$. This is not a null for want of power: the interval excludes effects
larger than about $0.016$, a quarter of the interaction measured in
Section~\ref{subsec:int-results}. Total variation agrees
($0.2345$ against $0.2310$, $p=0.53$), and the run-to-run spread does not
differ between the arms (Levene $p=0.74$).

One entry of Table~\ref{tab:angle_contrasts} is suggestive and should be
resisted. The convention~$\times$~ladder interaction is
$\genAngInterEst$ $[\genAngInterLo,\genAngInterHi]$ with $p=\genAngInterP$,
and the four cells of Table~\ref{tab:angle_cells} fall into a conspicuously
antisymmetric pattern, the two higher values lying on one diagonal. Across the
five contrasts of that table nothing survives Holm correction --- the smallest
adjusted $p$ is $0.30$ --- and with one target, one session and $\genAngN$
repetitions we read the pattern as coincidence. It would take its own campaign
to become anything else.

What this establishes is narrow: on this target, in this session, commanding a
negative rotation angle in one representative rather than the other produced no
detectable change in the prepared distribution, with the convention~$\times$~ladder
interval still admitting a shift of the A/B contrast of up to
$\genAngInterLo$. It is not a measurement of the pulse layer, for the reason
given above, and it does not retire the angle representative as a contributor
to the April--October difference. What it does do is bound that contribution: whatever the representative
contributes to the April--October difference, it is not a large mean shift on
this target. The remaining confounder, the tail, is the one
Section~\ref{sec:intervention} measures directly.

\subsection{The same circuit, two days apart}
\label{subsec:long-twoday}

The direct arm of that factorial is, gate for gate, the October Control
circuit: the instrument's records of the two acquisitions agree on all
$\genCmpCtlGates$ gates, their order, their controls and their angles, in both
ladders. The two were acquired on the same unit two days apart. That makes the
first two rows of Table~\ref{tab:compilations} a repeat of one circuit with
nothing whatever changed except the date, which is the comparison
Section~\ref{subsec:long-scope} had to do without.

The instrument moved, and by a large amount. Pooled fidelity rose from
$\genTwoDayFiveOct$ to $\genTwoDaySevenOct$, a shift of $\genTwoDayShift$
($p=\genTwoDayPooledP$), with $\genTwoDayShiftA$ on ladder~A
($p=\genTwoDayPA$) and $\genTwoDayShiftB$ on ladder~B
($p=\genTwoDayPB$). Forty-eight hours changed the absolute performance of this
preparation by more than half the range spanned by every compilation in
Table~\ref{tab:compilations}.

The ladder contrast moved too, from $\genCmpCtlAB$ to $\genCmpDirAB$, but that
movement is not established: the difference of effects is
$\genTwoDayDiffEst$ $[\genTwoDayDiffLo,\genTwoDayDiffHi]$ with
$z=\genTwoDayDiffZ$ and $p=\genTwoDayDiffP$. That the operating state of a unit
changes enough over two days to move absolute fidelity by $\genTwoDayShift$ is
measured here, not assumed; that it carries the ladder contrast with it is
not.

\subsection{What the compilations show, and what the dates do not}
\label{subsec:long-scope}

Table~\ref{tab:compilations} is the central observation of this section. On one
target, with one logical circuit, the ladder contrast runs from $\genCmpCtlAB$
through zero to $\genCmpRedAB$. It changes sign, and the compilations it runs
across are equal as operators.

Three of its five rows carry the weight. The Control and Reduced rows come from
the same session, interleaved and paired, and the difference between them is
the interaction of Section~\ref{subsec:int-results}, $\genInterEst$
$[\genInterLo,\genInterHi]$: a single effect, estimated under
a counterbalanced acquisition order, that spans most of the table on its own. The direct and folded
rows come from one later session and differ by an angle representative for
which Section~\ref{subsec:long-angle} measures no mean effect. And the April row sits
between them at a gate count between them.

The April--October comparison does not support a claim about the operating
state of the instrument, because the two acquisitions differ in circuit as well
as in time. How much the choice of comparator matters is worth setting out.
Testing the April contrast against October Control gives a
difference of effects $\genDiffCtlEst$ $[\genDiffCtlLo,\genDiffCtlHi]$,
$z=\genDiffCtlZ$, $p=\genDiffCtlP$. Testing it against October Reduced gives
$\genDiffRedEst$ $[\genDiffRedLo,\genDiffRedHi]$, $z=\genDiffRedZ$,
$p=\genDiffRedP$. The first comparator differs from April by the tail, which
Section~\ref{subsec:int-schedule} identifies as the gate pair whose scheduling
is ladder-dependent; the second differs from it by a single zero-angle rotation
occupying the same position in both ladders. A longitudinal claim resting on the
first rests on the more confounded of the two, and the second does not support
one.

So the statement this section makes is about compilations, and the statement it
declines to make is about time. The rows of Table~\ref{tab:compilations} differ
in circuit \emph{and} in session, and only the two October~5 rows are free of
that. The parsimonious reading is that the contrast moves because the
compilations differ, since one measured interaction accounts for most of the
range; but Section~\ref{subsec:long-twoday} shows that the instrument alone can
move absolute performance by $\genTwoDayShift$ in two days, so parsimony here is
a judgement and not a demonstration.

What would settle it is now a short experiment rather than a conjecture. The
two-day repeat shows that a circuit-exact longitudinal comparison is
straightforward to acquire on this unit; what it lacks is a documented
recalibration between the two acquisitions and enough repetitions to resolve a
contrast difference of the size seen here. Section~\ref{sec:discussion} sets out
that design.

\subsection{Other differences between the acquisitions}
\label{subsec:long-caveats}

Three further differences between the April and October campaigns must be
stated rather than assumed harmless.

\paragraph{Acquisition order.} The October campaigns interleave the ladders run
by run. The April campaign did not: within the D1 session the FULL-stage runs
were acquired in alternating blocks of five, so ladder~B occupies a mean
position $5$ slots later than ladder~A. If the device drifted during that
session, the offset would bias the April contrast. Its size can be estimated, with care. Fidelity within that session rises with
run index at $+2.3\times10^{-4}$ per run, but the trend is not significant
($\rho=+0.20$, $p=0.17$), so the figure is a point estimate and not a bound.
Taken at face value the $5$-slot offset moves $A-B$ by about $-0.001$: ladder~B
occupies the later slots, so a rising trend favours~B and biases $A-B$
downwards, by roughly two per cent of the range in
Table~\ref{tab:compilations}. That is the direction that would make the April
contrast look \emph{less} positive, so the blocking could in principle have
contributed to the null it produced, though at a tenth of the size needed to
account for it.

\paragraph{Shot count.} The two campaigns were configured for different
nominal shot counts, $4096$ in April against $2048$ in October, and neither
figure is recorded in the instrument export. This is not a difference in
statistical precision: the parameter does not count independent projective
measurements on this readout (Section~\ref{subsec:acquisition}). What it does
mark is an undocumented difference in how the two acquisitions were
configured, and we record it as such.

\paragraph{Sample size.} Every acquisition in Table~\ref{tab:compilations}
provides $\genNreps$ runs per ladder on this target, so no comparison here is
limited by one of them being smaller.

\section{Intervention on the physical control layer}
\label{sec:pulse}

The preceding campaigns examined how changes to the submitted gate list affect
state preparation without deliberately changing the device's pulse
calibration. Here we complement those comparisons with a pulse-level study on
the Shenzhen unit. Because this study was conducted on a different device, we
analyse its within-unit comparisons on their own terms and use the gate-level
results only as context.

\subsection{D1 protocol and analysis}
\label{subsec:pulse-scope}

We studied FULL-stage preparation of D1 under three configurations, with five
executions labelled A and five labelled B in each. Control uses the compiled
circuit with the standard device calibration. Experiment~1 removes
identity-equivalent gates while retaining that calibration. Experiment~2
implements the preparation with a globally optimised physical pulse. In
Control and Experiment~1, A and B denote distinct Gray-code ladder
compilations. In Experiment~2, the same physical pulse is used for both
labels, so A and B retain the nominal ladder labels rather than denote
separate pulse-level implementations.

For each of the 30 executions, we analyse the measured eight-state probability
distribution. We calculate classical fidelity against the D1 target for each
execution, then report the mean and sample standard deviation of the five
run-level fidelities in each configuration--ladder cell. Mean distributions
are used to compare the A and B outputs and to calculate their single-qubit
marginals. Supplement~\supref{sup:pulse} gives the corresponding run-level
results.

We obtain the single-qubit marginals from these same eight-state
distributions. With computational basis states ordered as $q_0q_1q_2$, so that
$k=4q_0+2q_1+q_2$, the Experiment~2 marginals are
\[
\bigl(P(q_0{=}1),P(q_1{=}1),P(q_2{=}1)\bigr)
=
\begin{cases}
(0.507,\,0.561,\,0.384), & \text{nominal ladder A},\\
(0.526,\,0.565,\,0.374), & \text{nominal ladder B}.
\end{cases}
\]
The values show that the three qubits do not respond uniformly: in particular,
$P(q_2{=}1)$ remains below $0.4$ in both nominal ladder groups. Pulse-level
preparation
therefore does not restore all single-qubit marginals to one half.

\subsection{D1 results}
\label{subsec:pulse-results}

\begin{table}[htbp]
\centering\small
\caption{D1 FULL-stage results under the three pulse-level configurations.
Fidelity is calculated for each execution and reported as the mean $\pm$
sample standard deviation over five runs. A and B represent distinct ladder
compilations in Control and Experiment~1, but nominal ladder groups using the
same physical pulse in Experiment~2. The A/B fidelity difference uses the
run-level means; total variation compares the mean distributions. Differences
were calculated before rounding.}
\label{tab:pulse_ab}
\begin{tabular}{lcccc}
\toprule
Configuration & $F$, A & $F$, B & $F_A-F_B$ & $\TV(A,B)$ \\
\midrule
Control       & $0.9075\pm0.0846$ & $0.9380\pm0.0165$ & $-0.0304$ & 0.1630 \\
Experiment~1  & $0.9631\pm0.0171$ & $0.8951\pm0.0774$ & $+0.0680$ & 0.0721 \\
Experiment~2  & $0.9724\pm0.0061$ & $0.9588\pm0.0263$ & $+0.0137$ & 0.0315 \\
\bottomrule
\end{tabular}

\end{table}

For the two configurations that retain distinct A and B ladder compilations,
removing identity-equivalent gates reduces the total-variation distance
between the mean outputs from $0.1630$ to $0.0721$.
The difference between their mean run-level target fidelities changes from
$-0.0304$ to $+0.0680$. Identity pruning thus brings the two
measured distributions closer without making their preparation accuracies
more alike.

With the common physical pulse of Experiment~2, the two nominal ladder groups
have a total-variation distance of $0.0315$ and differ in
mean run-level target fidelity by $0.0137$. Their mean output distributions
are therefore close under one implementation. This does not constitute a
test of whether pulse optimisation eliminates the ladder dependence seen when
two distinct compilations are executed. The corresponding run-level results
are given in Supplement~\supref{sup:pulse}.

\subsection{Direct pulse-level preparation across the benchmark suite}
\label{subsec:pulse-suite}

To assess direct pulse-level preparation across the benchmark suite, we
performed 25 executions for each nominal ladder label A and B for every
target D0--D6, for a total of 350 executions. For a given target, both
nominal ladder groups used the same distribution-specific physical pulse.
We therefore pool their 50
run-level outcomes to characterise preparation accuracy and
between-execution variation. The $\pm$ values in
Table~\ref{tab:pulse_suite} are sample standard deviations across those
executions.

\begin{table}[htbp]
\centering\small
\caption{Direct pulse-level FULL-stage preparation on the Shenzhen unit.
For each target, two 25-execution nominal ladder groups using the same
target-specific pulse are pooled. Fidelity and total-variation distance (TV)
are calculated for
each execution and reported as mean $\pm$ sample standard deviation;
$F_{\max}$ is the largest run-level fidelity. Gate-level fidelities from the
other unit are shown for context, not as a controlled comparison of
preparation methods.}
\label{tab:pulse_suite}
\begin{tabular}{lccccc}
\toprule
Target & $n$ & Mean $F$ & Mean $\TV$ & $F_{\max}$
& Gate-level $F$ \\
\midrule
D0 & 50 & $0.937 \pm 0.057$ & $0.200 \pm 0.062$ & 0.989 & 0.962 \\
D1 & 50 & $0.954 \pm 0.063$ & $0.163 \pm 0.067$ & 0.992 & 0.909 \\
D2 & 50 & $0.953 \pm 0.013$ & $0.163 \pm 0.027$ & 0.973 & 0.953 \\
D3 & 50 & $0.912 \pm 0.026$ & $0.245 \pm 0.048$ & 0.990 & 0.871 \\
D4 & 50 & $0.969 \pm 0.015$ & $0.141 \pm 0.030$ & 0.989 & 0.882 \\
D5 & 50 & $0.917 \pm 0.021$ & $0.159 \pm 0.025$ & 0.956 & 0.943 \\
D6 & 50 & $0.967 \pm 0.012$ & $0.116 \pm 0.022$ & 0.989 & 0.964 \\
\bottomrule
\end{tabular}

\end{table}

Pooled mean fidelities range from $0.912$ for D3 to $0.969$ for D4. Across
all seven targets, the direct-pulse implementation attains mean fidelity above
$0.9$, although the degree of agreement with the target varies by
distribution. The gate-level values in Table~\ref{tab:pulse_suite} provide a
reference for the same benchmark family, but differences between the two
campaigns cannot be assigned to the preparation method alone because the
device and acquisition period also differ.

\subsection{Interpretation}
\label{subsec:pulse-scopestatement}

The D1 study shows that identity pruning and a common pulse-level
implementation produce different patterns of agreement between A and B. The
benchmark-wide study shows that direct pulse-level preparation can achieve
high classical fidelity for each target on the Shenzhen unit. Together, these
observations motivate closer study of physical control as a contributor to
preparation accuracy. They do not, however, isolate pulse calibration as the
cause of the gate-level effects in
Sections~\ref{sec:suite}--\ref{sec:compilations},
nor do the measured populations determine the state-level information
considered in Section~\ref{sec:density-record}.

\section{State-level information in the instrument record}
\label{sec:density-record}

Everything reported so far is a statement about measured output
distributions. These constrain the correlations of the prepared object and
exclude the simplest noise description (Section~\ref{subsec:correlators}), but
they say nothing about its coherences: whether the two realisations differ as
\emph{states} requires information the counts do not contain.

The instrument record appears to contain it. The SpinQuasar export carries, for
every job, an $8\times8$ complex matrix in an \texttt{execution\_matrix} field
alongside the population vector. We describe that object because a reader of the
deposit will find it. It is not used in any claim of this paper, and
Section~\ref{subsec:density-scope} says why.

\subsection{The exported object}
\label{subsec:density}

\paragraph{Provenance.} All $\genDMn$ runs of the October campaign are
covered: the seven duplicated job names of Section~\ref{subsec:provenance} are
resolved by matching the deposited populations, which picks out one export job
in each case. Over the full set the export reproduces our run-level records to
$\genDMpopDev$.

\paragraph{What the object is.} Each matrix is Hermitian to $\genDMherm$, has
unit trace to $\genDMtrace$, has smallest eigenvalue $\genDMminev$, and has as
its diagonal exactly the reported populations. It is therefore a valid density
matrix to numerical precision, and it is consistent with the measured counts by
construction. Its off-diagonal entries are not determined by those counts: among
the $\genDMn$ matrices, the two whose population vectors are closest --- total
variation $\genDMcohTV$ --- differ by up to $\genDMcohDiff$ in the magnitude of
their off-diagonal entries. Every experimental matrix is rank-deficient --- rank $3$ or $4$
($\genDMrankThree$ and $\genDMrankFour$ respectively, with $\genDMrankOther$
exceptions), with a median purity of $\genDMpurity$ --- while the simulation
matrices supplied in the same records are rank $\genDMsimRank$ throughout.
Section~\ref{subsec:density-scope} examines that rank deficiency over the whole
instrument record, where it is sharper than on this campaign alone.

\paragraph{What it would say if taken at face value.} Read as a reconstruction
of the prepared state, these matrices give a median overlap
$\langle\psi|\rho|\psi\rangle=\genDMstateF$ with the ideal target state, ranging
over $[\genDMstateFlo,\genDMstateFhi]$, against a median classical fidelity of
$\genDMclF$ on the same runs; the two quantities are positively but not tightly
associated (Spearman $\rho=\genDMspearman$). The ladder contrasts do not
transfer intact. In the Control configuration the two realisations are
indistinguishable at this level ($\genDMCtlA$ against $\genDMCtlB$,
$p=\genDMpCtl$), although at distribution level they differ there by the largest
margin in the campaign; in the Reduced configuration they do differ
($\genDMRedA$ against $\genDMRedB$, $p=\genDMpRed$), with ladder~B above
ladder~A as at distribution level. Were these numbers admissible, they would say that the compilation effect of
Section~\ref{sec:intervention} is not visible in the overlap with the target in
the configuration where the distribution-level effect is largest. Even then
they would not say that the two realisations prepare the same state: distinct
states can share an overlap with a given target, so a null in this quantity is
not a null at the state level.

\subsection{Why this is not reported as a result}
\label{subsec:density-scope}

The numbers of the preceding paragraph are not claims of this paper, and we
would ask a reader not to cite them as such. The reason is that the export does
not say where the matrices come from, and the matrices themselves do not
settle it.

Their structure is at least uniform. The export carries an
\texttt{execution\_matrix} for every job, so the audit can be run over all
$\genDMrecN$ Grover--Rudolph jobs the two units have produced, spanning
\genDMrecSpan{}. Their ranks are \genDMrecRanks{}
(\genDMrecRankHist{}): never~$1$, never full. Of the $\genDMrecTotalEv$
eigenvalues this yields, the vanishing ones do not merely take small values,
they take machine zero --- the largest in magnitude over the entire record is
$\genDMrecZeroMax$, with median $\genDMrecZeroMedian$ --- while the smallest
surviving eigenvalue anywhere in the record is $\genDMrecNonzeroMin$. The two
groups are separated by $\genDMrecGap$ orders of magnitude, and only
$\genDMrecMid$ eigenvalues of $\genDMrecTotalEv$ fall anywhere between
$10^{-12}$ and $10^{-3}$. Over the same $\genDMrecDiagN$ jobs the diagonal
reproduces the deposited populations exactly, the largest discrepancy being
zero to the last bit.

We record this structure as a description and draw no inference from it.
Nothing in it discriminates between a reconstruction and a construction: a
maximum-likelihood estimator constrained to the positive cone returns
rank-deficient states with exactly vanishing eigenvalues as a matter of course,
and the behaviour is well documented \cite{ScholtenBlumeKohout2018}.

The provenance is documented, and settles both questions at once. On enquiry,
the manufacturer specifies the following. For a three-qubit reconstruction the
server processes seven NMR readout settings, combines the resulting
coefficients over the $64$ Pauli products, and applies a positivity correction
before export. The coefficients come from the corresponding job's own readout,
although the software also uses simulated expectation values when selecting
among repeated coefficient estimates. The exported object is a normalised
reconstruction, not an unprocessed thermal-ensemble density matrix.

Two consequences follow, and they point in opposite directions. The first is
that the rank structure above is accounted for: it is produced by the
positivity correction and is not a property of the underlying physical
ensemble, so the uniformity we documented across the record is an artefact of
the processing and nothing more. The second is why these matrices still cannot
answer the question this section opened with. The quantity we would want them
for is a comparison between two compilations of the same unitary. An estimator
that consults simulated expectation values --- values computed from the ideal
circuit, which is identical for the two compilations --- is not independent of
the hypothesis under test. However small that intervention is in practice, it
is in the wrong place: it enters exactly where the two arms are supposed to
differ. A state-level contrast obtained this way could be a property of the
prepared states or a property of the selection rule, and the deposited file
does not separate them.

We therefore use these matrices for nothing, and record what they are so that
others need not ask again.

What would settle the matter is an independent tomography, under a protocol we
control, of a small number of runs in the Control and Reduced configurations of
both ladders. Few runs would be needed, because the distribution-level effect
such a measurement would have to reproduce is large. The documentary route is
closed: the specification above is the answer, and the answer is that the
$985+\genDMn$ matrices already in hand cannot serve this purpose, whatever
other purpose they serve.

Until that measurement exists, the claims of this paper are claims about
measured output distributions, and the correlators of
Section~\ref{subsec:correlators} are the furthest those distributions reach.

\section{Interpretation and limits}
\label{sec:discussion}

This section separates three registers: what the experiments observed, what we
offer as a hypothesis about why, and what remains open. The central observation is robust and was obtained under a design we control.
The mechanism behind it is not, and no longitudinal claim is made at all, the
instrument's own record having shown that the acquisitions a longitudinal claim
would compare differ in circuit as well as in time.

\subsection{What the experiments establish}
\label{subsec:disc-established}

\paragraph{Logically equivalent compilations are not experimentally
equivalent.} Ladders A and B implement the same unitary and command the same
angles, agreeing in the ideal model to $\genExactAB$
(Section~\ref{subsec:exactness}). Across the benchmark suite the FULL-stage
contrast between them is significant for $\genNsigHolm$ of seven targets under
Holm (Section~\ref{subsec:suite-ab}), and its sign is not fixed across targets.
The suite cannot say which property of a target makes the difference appear,
because target is aliased with session (Section~\ref{subsec:suite-limits}).

\paragraph{A unitary-preserving intervention acts differently on the two
realisations.} This is the strongest-design result in the paper. Removing gates whose joint action is
the identity --- leaving the unitary exactly unchanged while changing the
executed schedule --- produces a ladder\,$\times$\,intervention interaction of
$\genInterEst$ $[\genInterLo,\genInterHi]$, paired $p=\genInterP$, with the A/B
advantage reversing sign (Section~\ref{subsec:int-results}). The comparison is within one session, interleaved run by run, paired by
repetition, with every circuit verified against the analytic target before
acquisition. No session effect and no target effect can produce it, and the
pairing suppresses drift slower than the acquisition cadence without
cancelling it; Supplement~\supref{sup:intervention-unpaired} recomputes the
interval under three treatments that relax the independence assumption, and
the conclusion is unchanged under all three.

\paragraph{The difference is not a property of the target.} On one target,
with one logical circuit, the contrast runs from $\genCmpCtlAB$ to
$\genCmpRedAB$ across five acquisitions that are equal as operators
(Section~\ref{subsec:long-scope}). It changes sign. Two of the five come from
one session and are separated by the measured interaction above; two more come
from a later session and differ only in the representative of an angle, for
which Section~\ref{subsec:long-angle} measures no mean effect.

\paragraph{A reproduced output distribution does not certify the prepared
object.} The measured correlators are inconsistent with any global depolarising
description of the errors: a single contraction factor across the suite leaves a
residual of $\genCorrResidMax$, per-target factors range from $\genCorrLamMin$
to $\genCorrLamMax$, and on the uniform target --- which such a channel would
reproduce exactly --- a two-body correlation of $\genCorrDzeroZZ$ appears at a
classical fidelity of $\genCorrDzeroF$ (Section~\ref{subsec:correlators}).
Structure is redistributed, not merely damped.

\paragraph{One compilation choice produced no detectable effect on mean
fidelity.} Commanding the two level-1 inverse rotations as $720\dg-\theta$
rather than $-\theta$ --- the
same gate list, the same count, the same controls --- moves the mean classical
fidelity by $-0.0006$, with a $95\%$ interval of $[-0.0157,+0.0146]$
(Section~\ref{subsec:long-angle}). Three qualifications travel with that null
and we state them rather than leave them to be inferred. The instrument record
shows what was commanded, not what was delivered, so the two arms are known to
differ in the command and only presumed to differ in the pulse; if the pulse
differs in duration then the execution times differ too, and the schedule is
not held fixed after all. The interval is on the mean fidelity, which is one
functional of the law and does not establish that the distributions coincide.
And the convention~$\times$~ladder interval still admits a shift of the A/B
contrast as large as $\genAngInterLo$, so the result does not retire the
representative as a contributor to the A/B difference. What it does say is
that, on this target in this session, this choice produced no detectable mean
effect, which is more than the commanded numbers would suggest.

\paragraph{The operating state of a unit moves on a scale of days.} The October
Control circuit, reacquired gate for gate two days later on the same unit,
gives a pooled fidelity $\genTwoDayShift$ higher ($p=\genTwoDayPooledP$,
Section~\ref{subsec:long-twoday}). Whether the ladder contrast moves with it is
not established ($p=\genTwoDayDiffP$).

\paragraph{What is not established.} Three things, so that they are not
inferred from the above. We do not establish that the ladder contrast varies
with the operating state of the instrument over time: the comparison that would
have shown it is confounded with a compilation difference
(Section~\ref{subsec:long-scope}). We do not establish anything at the level of
the prepared state, and Section~\ref{subsec:density-scope} explains why the
matrices in the instrument record cannot stand in for that. And we do not
identify a mechanism, which is the subject of the next subsection.

\subsection{Mechanism: hypothesis, not identification}
\label{subsec:disc-mechanism}

A mechanism adequate to these data would have to account for four things at
once: the magnitude of the FULL-stage discrepancy; its variation across targets;
its variation across compilations of one fixed unitary; and the appearance of
correlations the target does not contain. We can say with some confidence what
does \emph{not} account for them.

A global depolarising description fails on the last two points and is excluded
quantitatively in Section~\ref{subsec:correlators}; we use that conclusion here
rather than re-arguing it.

An angle-perturbation description is the natural next candidate, and the
companion paper supplies the bound that goes with it: perturbing each of the
construction's nodal angles by at most $\eta$ radians changes the output law by
at most $\min(1,n\eta)$ in total variation
\cite{FalcoFalcoPomaresMatthies2026}. The factor is $n$ and not the number of
nodal angles because the amplitude of any one basis state depends on exactly
one angle per level, so the perturbations that reach it number $n$
(Supplement~\supref{sup:theory}). The bound is on the angles of the
construction; a compiled circuit executes more rotations than that, and the
model below perturbs every one of them. The model is
\begin{equation}
\Ry(\phi_k)\;\longrightarrow\;\Ry(\phi_k+\delta_k),\qquad
\delta_k\overset{\mathrm{iid}}{\sim}\mathcal{N}(0,\sigma_\phi^{2}),
\label{eq:angle-noise}
\end{equation}
and the $L0$ stage offers the one place in this campaign where $\delta$ can be
read off rather than fitted: a single rotation on a single qubit, whose realised
angle follows from inverting $P(q_0{=}1)=\sin^{2}(\phi/2)$ run by run. Doing so
on the raw marginals gives a root-mean-square deviation of
$\genAngleRms\dg$ with a bootstrap interval of
$[\genAngleRmsLo\dg,\genAngleRmsHi\dg]$.

We report that number with three reservations, and together they are why
Equation~\eqref{eq:angle-noise} stays a phenomenological hypothesis in this
paper rather than becoming a calibrated error budget.

First, it is not a calibration constant but an artefact of a processing choice.
An earlier analysis of the same runs, applying an affine readout correction to
the $q_0$ marginals before inverting, obtained a scale roughly half as large.
Section~\ref{subsec:conventions} gives the experimental reason we no longer
apply such a correction anywhere in this paper; a quantity that halves under a
post-processing step we have evidence to reject is not something to build a
budget on.

Second, it is one gate. $L0$ commands a single $\Ry$ on $q_0$ with no CNOT
anywhere in the circuit. Nothing licenses extrapolating its realised error to
the eleven rotations and eight CNOTs of the FULL circuit, and an envelope
computed that way is an illustration rather than a measurement.

Third, and most seriously, the model's own assumption fails at the one gate
where it can be tested. Under independent zero-mean perturbations the seven
session means of the inferred deviation should scatter about zero with a
standard deviation of about $\genAnglePredicted\dg$, which is the pooled
within-session spread of $\genAngleWithin\dg$ divided by $\sqrt{\genNreps}$.
They scatter with a standard deviation of $\genAngleBetween\dg$, more than ten
times that, ranging from $\genAngleMeanLo\dg$ to $\genAngleMeanHi\dg$, and they
differ across sessions at $p=\genAngleKW$. The angle error at this gate is
dominated by a systematic offset that changes between sessions, not by
independent fluctuation from pulse to pulse --- which is consistent with the
session-dependent instrument state measured, on a different observable, in
Section~\ref{subsec:stages}.

The model also fails to predict the effect. The two ladders command identical
angle multisets, but that does not by itself make their perturbed output laws
equal: a perturbation drawn at a given position propagates through the CNOT
pattern that follows it, and the ladders differ in exactly that pattern. The
question is therefore quantitative, and we settled it by simulation.

We evolved the four cells of the October campaign --- both ladders, Control and
Reduced, with the tail placed as the instrument records it --- under
$\Ry(\phi)\to\Ry(\phi+\varepsilon)$ on every rotation, over
$20{,}000$ trials per cell at each of five perturbation scales from $1\dg$ to
$10\dg$ spanning the measured $\genAngleRms\dg$, and under two error
structures: $\varepsilon$ redrawn for every gate of every run, and one
$\varepsilon$ per gate shared across the four cells within a run. Three things follow. The model is not degenerate: run by run it does produce
an A/B difference, because the perturbation at a given position propagates
through the CNOT pattern that follows it. Its \emph{mean} A/B contrast is
nevertheless zero at every scale and under both structures,
$|\cdot|\le0.0002$ against the measured $+0.0395$, and so is the
ladder\,$\times$\,intervention interaction, $-0.0002$ against $+0.0625$ --- a
factor of some three hundred, with a Monte Carlo standard error of
$9\times10^{-5}$. And the run-to-run spread it predicts is too small for the
contrasts to be fluctuations of it: at the measured scale the standard
deviation of the per-run A/B difference is $0.0098$ when $\varepsilon$ is
redrawn for every gate of every run and $3.2\times10^{-4}$ when one
$\varepsilon$ per gate is shared across the cells of a run, against
$\genNreps$-run sample standard deviations of $0.047$ to $0.077$ in the
campaign. The model does reproduce the
direction of the reduction effect, $+0.0032$ where we measure $+0.0140$ and
$+0.0765$, but symmetrically in the two ladders, which is the one feature the
campaign is about. No single perturbation scale fits both ladders either:
ladder~A's measured fidelity corresponds to about $6\dg$, while ladder~B's lies
below the model's value at $10\dg$.

What this excludes is not a family of models but the one simulated: a Gaussian
angle perturbation on every executed rotation, under the two dependence
structures and the five scales above. Within that scope the exclusion is
clean --- the model predicts none of the three compilation effects reported
here --- and it rests on simulation rather than on an argument from symmetry.
A perturbation with a different distribution, correlated along the schedule,
or acting on something other than the commanded angle is untouched by it. The
script and its output are deposited.

The angle-representative result of Section~\ref{subsec:long-angle} points the
other way, with a weaker claim attached to it. There the two arms differ in the
representative commanded for the two level-1 inverse rotations, and the mean
classical fidelity shows no detectable change, to within $0.016$. We do not read that as a measurement of the pulse layer with the
schedule held fixed: Section~\ref{subsec:long-angle} sets out why the delivered
pulse is inferred rather than observed, and why a longer pulse would also move
the execution times. It is one null, on one target, in one session, and it
constrains the problem less than its symmetry with the intervention result
suggests.

The hypothesis we favour is that the order and timing of pulses matter. On a liquid-state NMR register the spins are permanently coupled
and the qubits evolve whether or not a gate is addressed to them: an idle qubit
is not idle. Section~\ref{subsec:stages} gives the measured form of this --- at
$L0$, qubits that receive no gate at all carry a spurious population of
$\genSpamQone$ and $\genSpamQtwo$, varying significantly across sessions ---
and Section~\ref{subsec:int-schedule} gives the most economical account of the
interaction in those terms, resting on where in the executed schedule the
removed gates sit and, in ladder~B, on the final CNOT that separates the two
tail pulses. We offer that as the most economical hypothesis consistent with
what we measured. It is not an identification, and the design of this study
cannot make it one: no experiment here varies the schedule while holding the
gate list fixed.

\subsection{Open questions and what would settle them}
\label{subsec:disc-open}

Three experiments would move this from characterisation towards mechanism. They
are listed in what we judge to be descending order of value per session of
instrument time.

\paragraph{(i) A documented recalibration.} The same A/B comparison before and
after a recalibration of one unit, targets and acquisition order interleaved,
with the compiled gate list held fixed and verified on both sides. Two things
learned since make this both easier and more pressing than it looked. A
circuit-exact repeat is straightforward to acquire --- we have one, two days
apart, in Section~\ref{subsec:long-twoday} --- so the design reduces to adding
a documented calibration event between the arms and enough repetitions to
resolve a contrast difference of the size seen there, which $\genNreps$ per
cell did not. And the requirement that the gate list be verified identical on
both sides, against the instrument's record rather than against our own, is the
lesson of Section~\ref{subsec:long-three}.

\paragraph{(ii) Permutations of one target.} Permute the outcome labels of a
single law: identical entropy, identical contrast, identical distance from the
uniform law, different organisation in the dyadic tree, with the assignment of
logical to physical qubits controlled. This separates the structure of the
preparation tree from the descriptors of the distribution, which the present
suite cannot do, and because all arms share one session it also breaks the
target--session alias of Section~\ref{subsec:suite-limits}.

\paragraph{(iii) Separating the removed gates.} The intervention of
Section~\ref{sec:intervention} removes the zero-angle rotation and the tail
together and measures only their joint effect. Three arms --- tail only,
zero-angle rotation only, both --- would attribute the interaction to one or the
other, and the tail is the component whose position differs between the two
ladders.

Beyond these, the state-level question of Section~\ref{sec:density-record} needs
no new hardware at all if the reconstruction behind the exported matrices can be
specified; failing that, an independent tomography of a small number of runs
under a protocol we control would decide it, and few runs would be needed
because the distribution-level effect such a measurement would have to reproduce
is large.

\subsection{What this means for a loading routine in practice}
\label{subsec:disc-practice}

We close with the consequence we would want a user of an exact state-preparation
routine to take away, stated at the scope the evidence supports.

The compilation of an exact preparation circuit is not a neutral step. Two
compilers presented with the same target, the same gate set and the same
construction can emit different gate lists, each exact in the ideal model, and
on the device used here those lists do not perform alike: the sign of a
comparison between two realisations reversed across three of them. This happened
to us, between two of our own campaigns six months apart, and we did not notice
it from our own records --- it was visible only in the gate lists the instrument
itself logged.

Two practical points follow. A reported loading fidelity is a property of a
compiled circuit, not of the algorithm named in the paper, and quoting one
without the executed gate list leaves a reader unable to reproduce it even in
principle. And a figure of merit computed on the output distribution is a weak
certificate: on this suite a classical fidelity above $0.9$ coexists with a
total-variation distance of $\genLenientTV$ and with a dominant two-body
correlation reproduced with the wrong sign. Where the prepared state is to be
consumed by a downstream routine sensitive to correlations rather than to
single-outcome probabilities, that distinction is not a technicality.

\section*{Data and code availability}
\label{sec:data-availability}

Everything this paper reports, with the single exception stated at the end of
this section, can be recomputed from deposited material without access to the
instrument or to the vendor SDK.

\paragraph{Run-level data.} Three hardware campaigns are deposited in full, one
row per run, carrying the measured and reference distributions over the eight
outcomes, the single-qubit marginals, the metrics, the gate census and the
circuit metadata: the $\genSessNtotal$-run benchmark campaign of
Section~\ref{sec:suite} (\texttt{runs\_flat\_v2.csv}, with its JSONL log and
per-group summaries); the $100$-run intervention campaign of
Section~\ref{sec:intervention}; and the $100$-run angle-representative campaign
of Section~\ref{subsec:long-angle}. The readout-mitigation session behind
Table~\ref{tab:readout} is deposited with its raw, mitigated and partially
mitigated logs and its $8\times8$ confusion matrix. The acquisition scripts that
produced the two October campaigns are deposited with them, as is the
bit-order calibration script of Section~\ref{subsec:conventions}.

\paragraph{The instrument's own record.} Two exports of the SpinQuasar job
database are deposited, taken after the campaigns of $5$ and $7$ October. They
are a second and independent record of the same executions, written by the
device rather than by our pipeline, and they carry the compiled gate list
actually executed, the measured populations and the $8\times8$ matrices of
Section~\ref{sec:density-record}. Their coverage is uneven and
Section~\ref{subsec:provenance} states exactly how: complete for both October
campaigns, and $498$ of $\genSessNtotal$ runs for the benchmark campaign, with
the gap established by circuit fingerprinting rather than assumed.

\paragraph{Verification.} A verification package accompanies the paper. It
contains a dependency-free implementation of the Grover--Rudolph angle map,
both Gray-code ladders and an exact state-vector simulator, and scripts that
confirm the exactness claim of Section~\ref{subsec:exactness} at machine
precision, check our records against the instrument's, and test the internal
consistency of the pulse-level tables. It needs neither hardware nor the vendor
SDK, so every claim here that does not require instrument access can be checked
directly. The campaign simulator samples a finite number of shots, so an
analytic simulator is necessary to verify exactness at machine precision; the
package provides one.

\paragraph{Generated tables and author-supplied summaries.} The numerical
tables for the CEU campaigns are generated from deposited run-level data by
the verification package. For those tables, regeneration into an empty
directory reproduces the submitted files byte for byte. The package also
emits macros for the principal estimates, intervals and $p$-values of those
campaigns; some secondary figures are read off the generated tables and typed.
The Shenzhen tables in Section~\ref{sec:pulse} and
Supplement~\supref{sup:pulse} are author-supplied summaries. Y.L. confirms
that their fidelities are means and sample standard deviations of per-execution
values; independent regeneration remains pending receipt of the underlying
records.

\paragraph{Repositories.}
\url{https://github.com/afalco/grover-rudolph-practical-implementation}
(compilation engine, acquisition scripts, campaign data, instrument exports);
\url{https://github.com/afalco/gr-triangulum-characterisation}
(execution pipeline and the distribution configurations with their checksums);
\url{https://github.com/afalco/gr-triangulum-verification}
(verification suite, benchmark run-level data, device-log extract, and the
table-generating script).

\paragraph{The Shenzhen material.} Section~\ref{sec:pulse} and
Supplement~\supref{sup:pulse} report Y.L.'s revised summaries of 30 D1
executions and 350 direct-pulse benchmark executions. Fidelity is calculated
per execution, with means and sample standard deviations over the stated
groups. The underlying probability arrays and execution scripts were not
included with the revised tables reviewed for this integration; their public
deposition and independent regeneration of the summaries remain to be verified.
Earlier consistency checks of averaged distributions do not validate the
revised per-execution statistics. No conclusion of Sections~\ref{sec:suite}
to~\ref{sec:compilations} depends on this campaign.

\section*{Declarations}

\paragraph{Funding.}
A.F, D.F.-P., and J.-C. L. acknowledge support under 
the Agreement between the Directorate-General for Innovation of
the Ministry of Innovation, Industry, Trade and Tourism of the Generalitat
Valenciana and the Universidad CEU Cardenal Herrera. Y.L. received no funding
for his contribution to this work.

\paragraph{Competing interests.}
Y.L. is affiliated with Shenzhen SpinQ Technology Co., Ltd., the manufacturer of
the devices characterised in this study. The hardware experiments reported here
were performed on two SpinQ Triangulum NMR quantum computers: the unit at
Shenzhen SpinQ Technology (Section~\ref{sec:pulse}) and the unit at Universidad
CEU Cardenal Herrera (Sections~\ref{sec:suite} to~\ref{sec:compilations}).
Y.L. declares that this affiliation and the use of SpinQ hardware may be
perceived as competing interests. The remaining authors declare no competing
interests.

\paragraph{Author contributions.}
A.F. and D.F.-P. developed the exact construction and the ancilla-free
compilation; A.F. designed the staged diagnostic protocol, the benchmark suite,
the intervention campaign and the angle-representative experiment, and executed
the latter two; J.-C.L. implemented the automation pipeline and the analysis
code; Y.L. designed and executed the pulse-level campaign and the hardware
calibration study; A.F. carried out the verification of both campaigns against
the instrument record; all authors contributed to the interpretation of the
results and to the writing of the manuscript.

\paragraph{Data availability.}
See the ``Data and code availability'' section above.

\paragraph{Ethics approval.}
Not applicable.


\clearpage
\setcounter{section}{0}
\setcounter{table}{0}
\setcounter{figure}{0}
\setcounter{equation}{0}
\renewcommand{\thesection}{S\arabic{section}}
\renewcommand{\thetable}{S\arabic{table}}
\renewcommand{\thefigure}{S\arabic{figure}}
\renewcommand{\theequation}{S\arabic{equation}}

\begin{center}
  {\Large\bfseries Supplementary Material}\\[4pt]
  {\normalsize Experimental characterisation of compilation-dependent
  Grover--Rudolph probability distribution loading on a three-qubit NMR
  processor}
\end{center}
\medskip

\section{The Grover--Rudolph construction and its compilation}
\label{sup:theory}

This section records the notation and the results of the companion paper
\cite{FalcoFalcoPomaresMatthies2026} on which the main text depends. It is a
dependency list, not a summary: the development and all proofs are in that
paper.

\subsection{The dyadic tree and the angle map}

A target law $p$ on $\{0,1\}^{n}$ is read as a measure on the leaves of a dyadic
tree of depth $n$. For a prefix $w\in\{0,1\}^{k}$ with $k<n$, write
$p_w=\sum_{x\succ w}p_x$ for the mass of the node reached by $w$, and $p_{w0}$,
$p_{w1}$ for the masses of its two children. Where $p_w>0$ the splitting angle
at that node is
\begin{equation}
\theta_w=\arccos\sqrt{p_{w0}/p_w}\in[0,\pi/2],
\label{eq:sup-angle-map}
\end{equation}
and the gate the hardware executes is $\Ry(\phi_w)$ with $\phi_w=2\theta_w$. A
node of zero mass contributes no constraint and its angle is set to zero by
convention; this is where the degenerate slots of
Section~\ref{sup:stages} come from.

Applying $\Ry(\phi_\emptyset)$ to $q_0$, then $\Ry(\phi_{w})$ to $q_k$
conditioned on the prefix $w$ held by $q_0\ldots q_{k-1}$, produces
\begin{equation}
\ket{\psi_p}=\sum_{x\in\{0,1\}^{n}}\sqrt{p_x}\,\ket{x},
\label{eq:sup-state}
\end{equation}
exactly. Two quantitative results of the companion paper are used in the main
text. Perturbing each of the construction's $2^{\,n}-1$ nodal angles by at most
$\eta$ radians changes the output law by at most $\min(1,n\eta)$ in total
variation, which is the envelope discussed in
Section~\mref{subsec:disc-mechanism} of the main text. The factor $n$ is the
number of \emph{levels}, not the number of angles: the nodal angles are
arranged in $n$ levels, with $2^{k}$ of them at level $k$, and the amplitude
$\sqrt{p_x}$ of a given basis state $x$ is a product of one factor per level,
each depending on the single angle at the node that $x$ passes through. Only
$n$ of the perturbations therefore reach any given amplitude, however many
angles the construction has in total. Two things the bound is not: it is not
stated for the elementary rotations a compilation executes, of which the FULL
circuit here has eleven against the $2^{3}-1=7$ nodal angles, and it is an
envelope over worst-case perturbations rather than a prediction for random
ones. And $S\ge
2^{\,n+1}\log(2/\delta)/\varepsilon^{2}$ repetitions suffice to estimate the
output law to total-variation accuracy $\varepsilon$ with confidence
$1-\delta$.

\subsection{The Gray-code ladder and the A/B freedom}

Level $k$ of the construction applies to $q_k$ a rotation whose angle depends on
the $k$-bit prefix already prepared: a uniformly controlled $\Ry$ with $k$
controls. Its ancilla-free decomposition \cite{Mottonen2005,Bergholm2005} writes
it as $2^{k}$ rotations interleaved with $2^{k}$ CNOTs,
\begin{equation}
\mathrm{UCR}_y(\phi_{0\ldots0},\dots,\phi_{1\ldots1})
=\prod_{j=1}^{2^{k}}\Ry(\alpha_j)\,\mathrm{CNOT}_{c_j\to q_k},
\label{eq:sup-ladder}
\end{equation}
where the coefficients $\alpha_j$ are a fixed linear transform of the
$\phi_w$ and the control sequence $(c_j)$ visits the $k$ control qubits in a
Gray-code order. For $k=2$ there are two such orders,
$(q_1,q_0,q_1,q_0)$ and $(q_0,q_1,q_0,q_1)$. They give the same unitary; they
are ladders~A and~B. Exchanging them permutes the coefficients $\alpha_j$, which
is why the two non-zero sub-rotations appear in the opposite order in the two
panels of Figure~\mref{fig:circuits} of the main text.

\begin{table}[htbp]
\caption{Commanded angles $\phi_w$ in degrees for the seven benchmark targets,
from Equation~\eqref{eq:sup-angle-map}. Subscripts index the dyadic prefix:
$\phi_\emptyset$ is the root rotation on $q_0$, $\phi_0$ and $\phi_1$ the level-1
rotations on $q_1$, and $\phi_{00}\ldots\phi_{11}$ the level-2 rotations on
$q_2$. These are the angles of the construction; the coefficients actually
emitted into the circuit are the ladder transform of the level-2 block.}
\label{tab:sup-angles}
\centering\small
\begin{tabular}{lccccccc}
\toprule
Target & $\phi_\emptyset$ & $\phi_0$ & $\phi_1$ & $\phi_{00}$ & $\phi_{01}$ & $\phi_{10}$ & $\phi_{11}$ \\
\midrule
D0 & $90.00$ & $90.00$ & $90.00$ & $90.00$ & $90.00$ & $90.00$ & $90.00$ \\
D1 & $90.00$ & $113.58$ & $66.42$ & $109.47$ & $98.21$ & $81.79$ & $70.53$ \\
D2 & $45.57$ & $65.68$ & $62.18$ & $80.41$ & $78.46$ & $74.17$ & $60.00$ \\
D3 & $90.00$ & $143.13$ & $36.87$ & $101.54$ & $141.06$ & $38.94$ & $78.46$ \\
D4 & $90.00$ & $136.05$ & $43.95$ & $115.38$ & $116.22$ & $63.78$ & $64.62$ \\
D5 & $59.04$ & $62.40$ & $131.53$ & $103.19$ & $167.63$ & $141.88$ & $79.59$ \\
D6 & $106.22$ & $125.78$ & $121.76$ & $67.85$ & $79.10$ & $78.34$ & $34.02$ \\
\bottomrule
\end{tabular}

\end{table}

\section{Compiled circuits for the truncated stages}
\label{sup:circuits}

\begin{figure}[htbp]
\centering
\begin{subfigure}[b]{\linewidth}
\centering
\begin{quantikz}[row sep=0.40cm, column sep=0.30cm]
\lstick{$q_0$} & \gate{\Ry(90\dg)} & \meter{} \\
\lstick{$q_1$} & \qw & \meter{} \\
\lstick{$q_2$} & \qw & \meter{}
\end{quantikz}
\caption{$L0$: the root rotation alone, $\genGatesLzeroNoTail$ gate. Qubits
$q_1$ and $q_2$ receive nothing, so their measured population has ideal value
zero for every target: the leakage observable of
Section~\ref{sup:stages-leakage}.}
\label{fig:sup-l0}
\end{subfigure}

\vspace{0.4cm}

\begin{subfigure}[b]{\linewidth}
\centering
\resizebox{\linewidth}{!}{%
\begin{quantikz}[row sep=0.40cm, column sep=0.20cm]
\lstick{$q_0$} &
  \gate{\Ry(90\dg)} & \gate{X} & \ctrl{1} & \qw & \ctrl{1} & \gate{X} &
  \qw & \ctrl{1} & \qw & \ctrl{1} & \meter{} \\
\lstick{$q_1$} &
  \gate{\Ry(56.79\dg)} & \qw & \targ{} & \gate{\Ry(663.21\dg)} & \targ{} & \qw &
  \gate{\Ry(33.21\dg)} & \targ{} & \gate{\Ry(686.79\dg)} & \targ{} & \meter{} \\
\lstick{$q_2$} &
  \qw & \qw & \qw & \qw & \qw & \qw & \qw & \qw & \qw & \qw & \meter{}
\end{quantikz}}
\caption{$L01$: levels $0$ and $1$, $\genGatesLoneNoTail$ gates. Only $q_2$ is
untouched.}
\label{fig:sup-l01}
\end{subfigure}
\caption{The truncated stages for~D1, ladder~A, transcribed from the gate lists
the instrument recorded on 1~April 2026. The FULL-stage panels are Figure~\mref{fig:circuits} of
the main text.}
\label{fig:sup-circuits}
\end{figure}
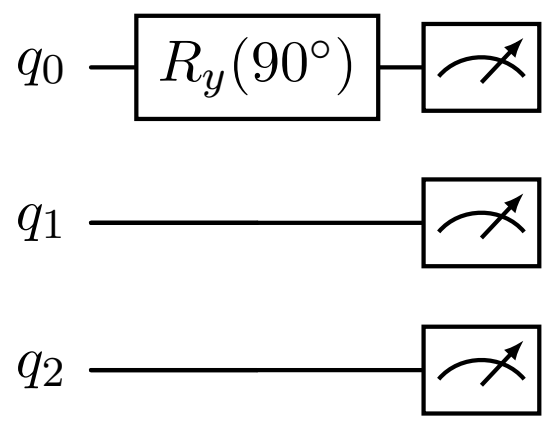
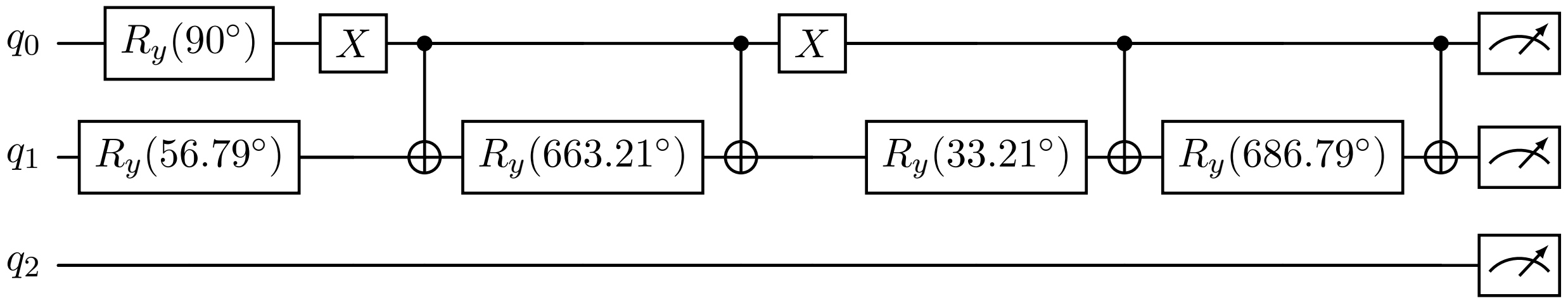

Figure~\ref{fig:sup-circuits} shows the two truncated stages as the instrument
executed them in the benchmark campaign. Neither stage carries the identity-safe tail --- the compiler of that campaign
did not emit it, which is why these circuits have $\genGatesLzeroNoTail$ and
$\genGatesLoneNoTail$ gates rather than the $\genGatesLzero$ and
$\genGatesLone$ of the compilation used in October. And the two level-1 inverse
rotations are commanded as $+663.211\dg$ and $+686.789\dg$ rather than as
$-56.789\dg$ and $-33.211\dg$: this is the angle convention discussed in
Section~\mref{subsec:long-three} of the main text, visible here in the circuit the instrument
recorded.

\section{Stage-wise analysis}
\label{sup:stages}

The main text uses the truncated stages for one purpose only
(Section~\mref{subsec:stages}) and states that they are not a depth series. This section
gives the analysis behind that restraint.

\subsection{Implementation error and incompleteness}
\label{sup:stages-decomposition}

A stage's distance from the final target mixes two things: how badly the stage
was executed, and how far that stage's own ideal output is from the target it has
not yet finished preparing. Separating them is straightforward. For stage $s$
with measured law $q_s$ and ideal law $p_s^{\mathrm{ideal}}$, write
\begin{equation}
E_s=\TV\!\left(q_s,\,p_s^{\mathrm{ideal}}\right),
\qquad
A_s=\TV\!\left(p_s^{\mathrm{ideal}},\,p^{*}\right),
\label{eq:sup-decomposition}
\end{equation}
for the implementation error and the incompleteness, with $p^{*}$ the final
target. By construction $A_{\mathrm{FULL}}=0$, so at the full stage the distance
to the target is the implementation error and nothing else.

\begin{table}[htbp]
\caption{The decomposition of Equation~\eqref{eq:sup-decomposition} for every
target, ladder~A. Entries are means over $\genNreps$ runs with sample standard
deviations; $A_s$ is a property of the ideal circuit and carries no error.}
\label{tab:sup-stages}
\centering\small
\begin{tabular}{llccc}
\toprule
Target & Stage & $E_s$ (implementation) & $A_s$ (incompleteness) & $\TV$ vs final target \\
\midrule
D0 & L0 & $0.101 \pm 0.037$ & $0.750$ & $0.649 \pm 0.037$ \\
   & L01 & $0.161 \pm 0.020$ & $0.500$ & $0.363 \pm 0.025$ \\
   & FULL & $0.163 \pm 0.030$ & $0.000$ & $0.163 \pm 0.030$ \\
\addlinespace
D1 & L0 & $0.099 \pm 0.039$ & $0.750$ & $0.651 \pm 0.039$ \\
   & L01 & $0.150 \pm 0.036$ & $0.500$ & $0.385 \pm 0.016$ \\
   & FULL & $0.268 \pm 0.031$ & $0.000$ & $0.268 \pm 0.031$ \\
\addlinespace
D2 & L0 & $0.108 \pm 0.019$ & $0.580$ & $0.477 \pm 0.019$ \\
   & L01 & $0.198 \pm 0.033$ & $0.400$ & $0.268 \pm 0.013$ \\
   & FULL & $0.140 \pm 0.026$ & $0.000$ & $0.140 \pm 0.026$ \\
\addlinespace
D3 & L0 & $0.079 \pm 0.024$ & $0.580$ & $0.502 \pm 0.021$ \\
   & L01 & $0.101 \pm 0.033$ & $0.500$ & $0.411 \pm 0.023$ \\
   & FULL & $0.335 \pm 0.056$ & $0.000$ & $0.335 \pm 0.056$ \\
\addlinespace
D4 & L0 & $0.129 \pm 0.040$ & $0.670$ & $0.541 \pm 0.040$ \\
   & L01 & $0.113 \pm 0.023$ & $0.500$ & $0.398 \pm 0.023$ \\
   & FULL & $0.290 \pm 0.056$ & $0.000$ & $0.290 \pm 0.056$ \\
\addlinespace
D5 & L0 & $0.117 \pm 0.021$ & $0.782$ & $0.694 \pm 0.017$ \\
   & L01 & $0.182 \pm 0.019$ & $0.660$ & $0.500 \pm 0.014$ \\
   & FULL & $0.115 \pm 0.031$ & $0.000$ & $0.115 \pm 0.031$ \\
\addlinespace
D6 & L0 & $0.123 \pm 0.041$ & $0.857$ & $0.745 \pm 0.031$ \\
   & L01 & $0.174 \pm 0.025$ & $0.241$ & $0.229 \pm 0.033$ \\
   & FULL & $0.198 \pm 0.034$ & $0.000$ & $0.198 \pm 0.034$ \\
\addlinespace
\bottomrule
\end{tabular}

\end{table}

Table~\ref{tab:sup-stages} does not show a depth curve. The implementation error
increases monotonically from $L0$ through $L01$ to FULL for
$\genStageMono$ of the seven targets only; for $\genStageBelowList$ the
FULL-stage implementation error is \emph{below} the $L0$ value, on a circuit
nineteen times longer.

A monotone curve would not be a depth effect either, which matters more. The
three stages do not differ only in depth:
each adds gates, changes the ideal state, and changes the support of the ideal
distribution, all at once. A comparison across them therefore confounds how hard
a circuit is to execute with how hard its target is to reach, and no arrangement
of these three points separates the two. The stages are reported as auxiliary
controls throughout, and the main text draws exactly one conclusion from them.

\subsection{Support leakage and spurious populations}
\label{sup:stages-leakage}

At the truncated stages the ideal state is supported on a subset of the eight
computational basis states, so mass measured outside that support has nowhere
legitimate to come from. Across the suite it runs from $\genLeakLzeroLo$ to
$\genLeakLzeroHi$ at $L0$ and from $\genLeakLoneLo$ to $\genLeakLoneHi$ at
$L01$.

Support leakage and the spurious population on the nominally inactive qubits are
the same quantity differently expressed, not two independent diagnostics: at
$L0$ the ideal support is exactly the set of strings with $q_1=q_2=0$, so mass
off the support is mass on those qubits. The main text reports the per-qubit form, in Table~\mref{tab:spam}, because
its ideal value is zero for every target. The measured value need not be: the
root rotation $L0$ applies is target-dependent, so a leakage channel driven by
it would carry that dependence to the idle qubits. This is why the comparison
below is restricted to the targets sharing a root angle.

We call the quantity an effective state-preparation-and-measurement error. The
word effective is doing work: these data do not separate imperfect preparation,
residual coupling between permanently coupled spins, and readout assignment
error, and nothing in this campaign could.

The quantity is \emph{not} constant across the campaign: it differs across the
seven sessions at $p=\genSpamKWqone$ and $p=\genSpamKWqtwo$ for the two qubits,
with the two-qubit mean ranging from $\genSpamRangeLo$ to $\genSpamRangeHi$.
Across all seven that figure mixes two things, since the root angle differs
between targets. The $\genLzeroSubsetN$ targets with a root angle of
$\genLzeroSubsetRoot\dg$ --- \genLzeroSubsetTargets{} --- execute the same
single gate at $L0$, and among them the leakage still runs from
$\genLzeroSubsetLo$ to $\genLzeroSubsetHi$, a factor of
$\genLzeroSubsetRatio$, at $p=\genLzeroSubsetKW$. With the circuit fixed, the
quantity is a property of the processor at a time and not of the processor.

\subsection{Why the stages do not predict FULL-stage performance}
\label{sup:stages-nopredict}

\begin{table}[htbp]
\caption{Rank correlation across the seven targets between each truncated-stage
diagnostic and the pooled FULL-stage classical fidelity. Seven independent
targets, however many runs each contains.}
\label{tab:sup-stage-predictors}
\centering\small
\begin{tabular}{lcc}
\toprule
Stage diagnostic & $\rho$ with FULL $F$ & $p$ \\
\midrule
$E_{L0}$ & $+0.32$ & $0.482$ \\
$E_{L01}$ & $+0.71$ & $0.071$ \\
$F_{L0}$ & $-0.32$ & $0.482$ \\
$F_{L01}$ & $-0.50$ & $0.253$ \\
Leakage $L0$ & $+0.32$ & $0.482$ \\
Leakage $L01$ & $+0.11$ & $0.819$ \\
\bottomrule
\end{tabular}

\end{table}

Does a cheap shallow measurement tell you what the expensive deep one will do?
On this suite it does not. No diagnostic reaches
significance, the largest correlation is $\genStagePredMaxRho$ at
$p=\genStagePredMaxP$, and that largest one has the wrong sign --- it associates
a \emph{larger} implementation error at $L01$ with a \emph{higher} FULL-stage
fidelity. With seven targets the test is weak, and we draw no conclusion beyond
the negative one: the staged protocol is a set of auxiliary controls and is used
as nothing else.

\section{Multiple comparisons across the benchmark suite}
\label{sup:multiplicity}

Section~\mref{subsec:suite-ab} of the main text tests one hypothesis seven
times, once per target: that the two Gray-code realisations perform identically
at the FULL stage. It reports the count of significant contrasts under Holm,
notes that Benjamini--Yekutieli agrees and that Benjamini--Hochberg gives one
more. This section sets out why those three procedures are the relevant ones,
why they disagree here, and why the Holm count is the one we lead with.

\subsection{Two error quantities}

With $m$ hypotheses of which $m_0$ are true, let $V$ be the number of true
hypotheses rejected and $R$ the total number rejected. Two quantities can be
controlled, and they are not the same thing:
\begin{equation}
\FWER=\PR(V\ge1),
\qquad
\FDR=\E\!\left[\frac{V}{R\vee1}\right].
\label{eq:sup-errors}
\end{equation}
The first is the probability of \emph{any} false rejection; the second is the
expected share of the rejections that are false. Since
$V/(R\vee1)\le\mathbf{1}\{V\ge1\}$ pointwise, $\FDR\le\FWER$ always, so a
procedure controlling the FWER controls the FDR but not conversely. FDR control
is the weaker guarantee and buys power with it.

Which one a paper should control depends on what a false positive costs. Here a
single spurious ladder contrast would be reported as evidence that two
logically equivalent compilations differ on hardware --- the paper's central
claim --- so we take the conservative side.

\subsection{The three procedures}

Order the raw $p$-values ascending, $p_{(1)}\le\cdots\le p_{(m)}$. All three
procedures are most conveniently written as adjusted $p$-values, compared
against $\alpha$ directly.

\paragraph{Holm \cite{Holm1979}} controls the FWER under \emph{any} dependence:
\begin{equation}
\tilde p_{(i)}^{\,\mathrm{Holm}}
=\max_{j\le i}\;\min\bigl\{(m-j+1)\,p_{(j)},\,1\bigr\}.
\label{eq:sup-holm}
\end{equation}
It is a step-down procedure and is uniformly more powerful than the Bonferroni
correction \cite{Dunn1961}, which multiplies every $p$-value by $m$.

\paragraph{Benjamini--Hochberg \cite{BenjaminiHochberg1995}} controls the FDR
at level $\alpha$ when the test statistics are independent, or positively
dependent in the PRDS sense:
\begin{equation}
\tilde p_{(i)}^{\,\mathrm{BH}}
=\min_{j\ge i}\;\min\bigl\{(m/j)\,p_{(j)},\,1\bigr\}.
\label{eq:sup-bh}
\end{equation}

\paragraph{Benjamini--Yekutieli \cite{BenjaminiYekutieli2001}} controls the FDR
under \emph{arbitrary} dependence, at the cost of a factor
$c(m)=\sum_{k=1}^{m}1/k$, which for $m=7$ is $\genBYconst$:
\begin{equation}
\tilde p_{(i)}^{\,\mathrm{BY}}
=\min_{j\ge i}\;\min\bigl\{c(m)\,(m/j)\,p_{(j)},\,1\bigr\}.
\label{eq:sup-by}
\end{equation}

\subsection{Applied to the seven contrasts}

\begin{table}[htbp]
\caption{The seven A/B contrasts at the FULL stage, ordered by raw $p$-value,
with the three adjustments. Raw values are two-sided Mann--Whitney
\cite{MannWhitney1947} over $\genNreps$ runs per cell. Bold marks an adjusted
$p$ below $0.05$.}
\label{tab:sup-multiplicity}
\centering\small
\begin{tabular}{clccccccc}
\toprule
$i$ & Target & $\bar F_A$ & $\bar F_B$ & $p_{(i)}$ & Holm & BH & BY \\
\midrule
1 & D6 & $0.950$ & $0.979$ & $<10^{-4}$ & $\mathbf{<10^{-4}}$ & $\mathbf{<10^{-4}}$ & $\mathbf{<10^{-4}}$ \\
2 & D0 & $0.954$ & $0.970$ & $0.0002$ & $\mathbf{0.0014}$ & $\mathbf{0.0008}$ & $\mathbf{0.0021}$ \\
3 & D5 & $0.951$ & $0.934$ & $0.0019$ & $\mathbf{0.0095}$ & $\mathbf{0.0044}$ & $\mathbf{0.012}$ \\
4 & D3 & $0.859$ & $0.883$ & $0.014$ & $0.058$ & $\mathbf{0.025}$ & $0.066$ \\
5 & D2 & $0.950$ & $0.957$ & $0.112$ & $0.335$ & $0.156$ & $0.405$ \\
6 & D1 & $0.908$ & $0.909$ & $0.801$ & $1.000$ & $0.846$ & $1.000$ \\
7 & D4 & $0.882$ & $0.882$ & $0.846$ & $1.000$ & $0.846$ & $1.000$ \\
\midrule
\multicolumn{5}{r}{rejected at $\alpha=0.05$} & $\mathbf{3}$ & $\mathbf{4}$ & $\mathbf{3}$ \\
\bottomrule
\end{tabular}

\end{table}

Exactly one target changes status between the procedures.
$\genMultFlipTarget$ has the $\genMultFlipRank$th smallest raw $p$-value,
$\genMultFlipRaw$. Holm multiplies it by $m-i+1=\genMultFlipHolmFactor$, giving
$\genMultFlipHolm$; BH multiplies it by $m/i=\genMultFlipBHFactor$, giving
$\genMultFlipBH$. The whole of the difference between ``four of seven'' and
``three of seven'' is that one factor.

\paragraph{Is the independence assumption defensible?} The seven targets were
not interleaved: each was acquired in its own contiguous session on one
instrument over three days (Table~\mref{tab:sessions}), and
Section~\ref{sup:stages-leakage} shows, on sessions executing an identical
$L0$ circuit, that the instrument was not in the same state throughout. Any such drift induces
dependence between the contrasts. The dependence is plausibly \emph{positive}
--- a session in which the machine performs poorly depresses both ladders of
that target together --- which is the PRDS case under which plain BH is valid.
But that is an argument from plausibility, not a property the design
guarantees, and it cannot be checked from seven points.

Benjamini--Yekutieli assumes nothing, costs the factor $\genBYconst$, and
returns the same three targets as Holm. That coincidence is the reason the main
text puts it as it does: \textbf{both procedures that make no independence
assumption select the same set, and the count of four arises only under the
procedure whose validity rests on an assumption this design does not
guarantee.}

\subsection{Cautions}

Three, recorded because they bear on how the numbers above should be read.

\emph{The procedure was fixed before the adjustment was computed.} Choosing
between Holm and BH after seeing which gives the more agreeable count would
invalidate both. We report all three and lead with the most conservative.

\emph{Adjusted $p$-values are not evidence strengths.} An adjusted $p$ of
$0.058$ does not mean the D3 contrast is absent; it means the evidence does not
survive a correction for having looked seven times. The estimate and its
interval, not the verdict, carry the information, and the main text reports
both.

\emph{The family is the seven targets of one campaign.} No correction in this
paper spans campaigns, because the intervention and angle-representative
campaigns each test a pre-specified contrast in a single pre-specified design
rather than searching a family. Where several contrasts are reported from one
of those campaigns --- Table~\mref{tab:angle_contrasts} --- the main text
applies Holm within that table and says so.

\section{Structural descriptors and the exploratory asymmetry analysis}
\label{sup:asymmetry}

Section~\mref{subsec:suite-asymmetry} of the main text reports an exploratory association between the
asymmetry of the dyadic tree and FULL-stage fidelity. This section gives the
descriptors and the full analysis.

Two descriptors capture tree asymmetry. The level-1 deviation
$\Delta_1=|\phi_0-\phi_1|$ measures how unequally the first refinement treats
the two halves of the mass. The UCR$_y$ range is the spread of the four level-2
angles, $\max_w\phi_w-\min_w\phi_w$ over $|w|=2$. Alongside them we take two
descriptors of the distribution itself, the Shannon entropy and the contrast
$p_{\max}/p_{\min}$, and the total-variation distance from the uniform law.

\begin{table}[htbp]
\caption{Structural descriptors of the seven targets, with the rank correlation
of each against pooled FULL-stage classical fidelity.}
\label{tab:sup-descriptors}
\centering\small
\begin{tabular}{lcccccc}
\toprule
Target & $\Delta_1$ (\dg) & UCR$_y$ range (\dg) & $H$ (bits) & Contrast & $\TV$ to uniform & FULL $F$ \\
\midrule
D0 & $0.00$ & $0.00$ & $3.00$ & $1.00$ & $0.00$ & $0.962$ \\
D1 & $23.58$ & $38.94$ & $2.85$ & $4.00$ & $0.20$ & $0.909$ \\
D2 & $27.82$ & $20.41$ & $2.45$ & $35.00$ & $0.38$ & $0.953$ \\
D3 & $53.13$ & $102.12$ & $2.02$ & $20.00$ & $0.55$ & $0.871$ \\
D4 & $46.05$ & $52.45$ & $2.44$ & $15.50$ & $0.37$ & $0.882$ \\
D5 & $41.53$ & $88.05$ & $2.36$ & $144.35$ & $0.38$ & $0.943$ \\
D6 & $35.78$ & $45.08$ & $2.41$ & $19.14$ & $0.37$ & $0.964$ \\
\midrule
$\rho$ with $F$ & $-0.64$ & $-0.64$ & $+0.39$ & $-0.11$ & $-0.54$ & \\
$p$ & $0.12$ & $0.12$ & $0.38$ & $0.82$ & $0.22$ & \\
\bottomrule
\end{tabular}

\end{table}

The ordering in Table~\ref{tab:sup-descriptors} is the content of the claim: the
asymmetry descriptors correlate with fidelity more strongly than entropy or
contrast do, $\genDescDeltaRho$ for $\Delta_1$ against $\genDescEntRho$ for
entropy and $\genDescContRho$ for contrast. A matched comparison points the same
way: among the four targets lying within $\genMatchSpan$ of one another in
distance from the uniform law --- $\genMatchList$ --- the fidelity still spans
$\genMatchFidLo$ to $\genMatchFidHi$, and the ordering follows tree asymmetry
rather than any of the distributional descriptors.

Four reasons this is reported as exploratory and not as a predictor, each
sufficient on its own.

There are seven independent targets, however many runs each contains, and
$p=\genDescDeltaP$ does not establish an association. The two asymmetry
descriptors are collinear at $\rho=\genDescCollinear$
($p=\genDescCollinearP$): they are two parametrisations of one feature of the
tree, not two pieces of evidence. The matched comparison does not break the
confound it appears to address, because its four targets were acquired in four
different sessions, and Section~\ref{sup:stages-leakage} shows the instrument
differed measurably between sessions. And the main text's own result is that
FULL-stage performance depends on the compilation at fixed target, which means
no descriptor of the target alone can be expected to organise it.

The defensible statement is the one the main text makes: this suite shows an
exploratory association between tree asymmetry and performance, and testing it
requires an interleaved acquisition in which target and session are not aliased
--- the permutation experiment of Section~\mref{subsec:disc-open}.

\section{Extended statistics for the intervention campaign}
\label{sup:intervention-stats}

\subsection{Unpaired tests and intervals}
\label{sup:intervention-unpaired}

\begin{table}[htbp]
\caption{Unpaired counterparts of the paired contrasts reported in Table~\mref{tab:october_contrasts} of
the main text. Welch intervals and $p$-values, with Mann--Whitney $p$-values
alongside.}
\label{tab:sup-unpaired}
\centering\small
\begin{tabular}{lcccc}
\toprule
Contrast & Estimate & 95\% CI (Welch) & $p$ (Welch) & $p$ (Mann--Whitney) \\
\midrule
Ladder A $-$ B, Control & $+0.0395$ & $[+0.0135,\,+0.0656]$ & $0.004$ & $0.004$ \\
Ladder A $-$ B, Reduced & $-0.0229$ & $[-0.0439,\,-0.0019]$ & $0.033$ & $0.050$ \\
Reduction effect, ladder~A & $+0.0140$ & $[-0.0144,\,+0.0424]$ & $0.325$ & $0.286$ \\
Reduction effect, ladder~B & $+0.0765$ & $[+0.0591,\,+0.0939]$ & $2.7\times10^{-11}$ & $2.1\times10^{-8}$ \\
\bottomrule
\end{tabular}

\end{table}

Table~\ref{tab:sup-unpaired} repeats the October contrasts without using the
pairing. Every estimate is identical, as it must be --- the paired and unpaired
estimators of a difference of means coincide --- and every interval is wider,
every $p$-value larger. The qualitative conclusions do not change: the Control
ladder contrast and the reduction effect on ladder~B survive, the reduction
effect on ladder~A does not, and the Reduced ladder contrast sits at the
conventional threshold either way.

The paired analysis is the one the main text reports, for a reason worth
stating. Two distinct assumptions are at stake and it is worth keeping them
apart. One is about a single observation: what fraction of a run's deviation is
sampling noise. The other is about the relation between runs: whether the
repetitions are statistically independent of one another. Neither implies the
other.

The first is settled, and not in the direction a shot count would suggest.
Triangulum reads out an NMR ensemble, and the nominal \texttt{shots} parameter
scales probabilities into counts rather than counting projective measurements
(Section~\mref{subsec:acquisition}). A run therefore has no multinomial
sampling scale of its own, and the only sampling variation available is the
spread between repeated executions --- which is the quantity every interval in
this paper is built from.

The unpaired intervals need both assumptions, treating the $\genNreps$ runs of
a cell as independent draws from a fixed law. The paired analysis discharges
the second within a repetition: comparing cells acquired under the same
conditions tolerates arbitrary dependence there. What it still needs is
that the \genNreps{} \emph{differences} be uncorrelated across repetitions
with a common variance, which is what makes $s/\sqrt{n}$ the standard error of
their mean. Exchangeability is not enough for that: an exchangeable sequence
with common correlation $\rho$ has $\operatorname{Var}(\bar d)=
\sigma^{2}[1+(n-1)\rho]/n$, so even a small positive $\rho$ inflates the true
variance substantially at $n=\genNreps$. A drift slow on the scale of one
repetition but not on the scale of the session induces exactly that: adjacent
measurements reduce its influence without cancelling it, since they are still
taken at different times. The acquisition order was counterbalanced rather
than randomised (Section~\ref{sup:intervention-schedule}), which protects
against a monotone trend but not against autocorrelation.

\paragraph{Tested rather than assumed.} We therefore examined the
\genNreps{} differences directly; the script and its output are deposited. No
autocorrelation at lags $1$ to $5$ exceeds the Bartlett band of $\pm0.392$ ---
the largest is $+0.34$, on the ladder-A reduction effect, whose Durbin--Watson
of $1.03$ is the one value worth noting --- and no Ljung--Box test rejects,
though with \genNreps{} points that is weak evidence of absence. Two series do
carry a significant trend in repetition index, the A/B contrast in Reduced
($p=0.003$) and the ladder-B reduction effect ($p=0.012$), so the instrument
was drifting within the session.

We then recomputed every interval under three treatments that relax the
independence assumption rather than dispense with assumptions altogether. Each
replaces it with a weaker one, and we state which.

A Newey--West HAC standard error admits correlation at lags inside a chosen
window, weighting them by a Bartlett kernel and treating anything beyond as
negligible; the window here is the usual $\lfloor4(n/100)^{2/9}\rfloor$, which
at $n=\genNreps$ is two lags. A moving-block bootstrap at block length $3$
resamples contiguous triples, so dependence shorter than a block is carried
into the resample and dependence spanning blocks is broken by it --- the block
length sets what the procedure preserves, and does not establish that nothing
remains beyond it. A block sign-flip permutation test flips the sign of whole
blocks of three; its reference distribution is exact when the joint law of the
blocks is invariant under those sign changes, which holds if the blocks are
independent and each is symmetric about the null mean. Marginal symmetry of
each difference is not sufficient on its own, and with $\genNreps$ points the
block independence is assumed rather than verified. Both resampling treatments
use $20{,}000$ replicates. Standard errors inflate by $13$ to $26$ per cent,
which is modest.

The intervals do move; the conclusions do not. The interaction is
$[+0.0269,+0.0980]$ under HAC and $[+0.0254,+0.0928]$ under the block
bootstrap, against $[\genInterLo,\genInterHi]$ paired, with a block sign-flip
$p$ of $0.004$: zero stays excluded under all three. The Control contrast and
the ladder-B reduction effect behave the same way. The single exception is the A/B contrast in the
Reduced configuration, whose block-bootstrap interval reaches $+0.0001$ and
whose block-flip $p$ is $0.050$ --- the same contrast that fails a Holm
correction across the five rows of
Table~\mref{tab:october_contrasts}. Section~\mref{subsec:int-results} reports
it as an estimate rather than as a finding, for both reasons.

\subsection{Acquisition schedule and the instrument record}
\label{sup:intervention-schedule}

\begin{table}[htbp]
\caption{Position of each cell in the October acquisition schedule. The four
cells were visited once per repetition, with the order reversed on alternate
repetitions.}
\label{tab:sup-schedule}
\centering\small
\begin{tabular}{llcc}
\toprule
Configuration & Ladder & Runs & Mean position in the schedule \\
\midrule
Control & A & 25 & $50.4$ \\
Control & B & 25 & $50.5$ \\
Reduced & A & 25 & $50.5$ \\
Reduced & B & 25 & $50.6$ \\
\bottomrule
\end{tabular}

\end{table}

The campaign ran for $\genSchedSpan$ hours with a median of
$\genSchedMedianGap$ minutes between runs. The interleaving worked as designed:
the four cells have mean positions between $\genSchedPosLo$ and
$\genSchedPosHi$ in a schedule of one hundred slots, so no cell occupies
systematically earlier or later time.

Two drift checks. Across the session, fidelity rises with run index at
$\genSchedDriftSlope$ per run, $\rho=\genSchedDriftRho$ with
$p=\genSchedDriftP$ --- no detectable trend, and in any case common to all four
cells by the interleaving. And the acquisition was interrupted once, for
$\genSchedGapMin$ minutes before run~$\genSchedGapAt$; mean fidelity was
$\genSchedGapBefore$ before the interruption and $\genSchedGapAfter$ after it,
$p=\genSchedGapP$.

\paragraph{The seven orphaned jobs.} The instrument export contains seven jobs
carrying campaign run names that are not part of this dataset. An earlier
attempt at the campaign was launched into a directory that already held a
simulator dry run; because the resume logic of that version keyed on run name
alone, it treated the simulator rows as complete and began acquisition at
repetition~3. It then lost its connection to the instrument after seven jobs and
recorded nothing. The campaign was reacquired from scratch into a clean
directory, and every one of its hundred runs matches the \emph{later} of the two
device jobs to floating-point identity. The seven are documented here and in the
data deposit because a reader cross-checking by job name would otherwise find an
unexplained duplicate. The acquisition script now refuses to continue when a log
holds runs acquired on a different backend, so this particular failure cannot
recur.


The revised treatment below was supplied by Y.L., who ran the campaign.
It specifies per-execution fidelity summaries, the bit order used for the
recalculated marginals, and the use of the same physical pulse for both nominal
ladder groups in Experiment~2. These clarifications supersede the earlier
averaged-distribution treatment. The underlying probability arrays and
execution scripts were not included with the revised tables reviewed for this
integration; independent verification and public deposition remain to be
confirmed in the data availability statement.

\section{Pulse-level campaign: detailed results}
\label{sup:pulse}

For the D1 study, each of the six configuration--ladder cells contains five
measured eight-state probability distributions. For a run-level distribution
$p_r$ and target $p^*$, we calculate the classical fidelity
\[
F_r=\left(\sum_{k=0}^{7}\sqrt{p_k^*p_{r,k}}\right)^2.
\]
Table~\ref{tab:sup-pulse-d1-runs} reports the arithmetic mean and sample
standard deviation of the five run-level fidelities in each cell. The
total-variation distance and single-qubit marginals are calculated from the
cell's mean eight-state distribution. This is the same run-level fidelity
convention used for the D1 comparison in Section~\mref{sec:pulse}.

\begin{table}[htbp]
\centering\small
\caption{D1 FULL-stage results from 30 run-level probability distributions.
Classical fidelity is calculated for each execution and reported as the mean
$\pm$ sample standard deviation over five executions per cell. Total-variation
distance (TV) and the three marginals are calculated from each cell's mean
eight-state distribution. A and B denote distinct ladder compilations in
Control and Experiment~1, but nominal ladder groups using the same physical
pulse in Experiment~2.}
\label{tab:sup-pulse-d1-runs}
\resizebox{\textwidth}{!}{%
\begin{tabular}{llcccccc}
\toprule
Configuration & Nominal ladder & $n$ & Mean $F$ & $\TV(p^*,\bar p)$
& $P(q_0{=}1)$ & $P(q_1{=}1)$ & $P(q_2{=}1)$ \\
\midrule
Control & A & 5 & $0.908 \pm 0.085$ & 0.174
& 0.494 & 0.554 & 0.392 \\
Control & B & 5 & $0.938 \pm 0.017$ & 0.195
& 0.592 & 0.452 & 0.521 \\
Experiment~1 & A & 5 & $0.963 \pm 0.017$ & 0.139
& 0.469 & 0.515 & 0.433 \\
Experiment~1 & B & 5 & $0.895 \pm 0.077$ & 0.194
& 0.413 & 0.523 & 0.408 \\
Experiment~2 & A & 5 & $0.972 \pm 0.006$ & 0.130
& 0.507 & 0.561 & 0.384 \\
Experiment~2 & B & 5 & $0.959 \pm 0.026$ & 0.132
& 0.526 & 0.565 & 0.374 \\
\bottomrule
\end{tabular}%
}

\end{table}

We use the bit order $q_0q_1q_2$, with $k=4q_0+2q_1+q_2$. Thus, for example,
\[
P(q_0{=}1)=p_{100}+p_{101}+p_{110}+p_{111};
\]
the other marginals sum the states whose corresponding bit equals one. For
Experiment~2, the resulting marginals are $(0.507,\,0.561,\,0.384)$ for
nominal ladder A and $(0.526,\,0.565,\,0.374)$ for nominal ladder B. Under
both nominal ladder labels,
$P(q_2{=}1)$ remains below $0.4$, despite the close agreement between their
full output distributions.

For the direct-pulse study, each of the seven targets has 25 executions for
each nominal ladder label A and B, giving 350 executions overall. Because the
same distribution-specific physical pulse is used in both nominal ladder
groups for a given target, the analysis in Section~\mref{sec:pulse} pools all
50 run-level
outcomes per target. The reported mean, sample standard deviation, and
maximum are calculated from the individual-run target fidelities. These
50-run summaries describe the pooled nominal ladder groups, rather than
either 25-run group separately.

The reported standard deviations quantify variation across recorded NMR
ensemble executions. They are not inferred from the nominal SDK shots
parameter, which is not treated here as a count of independent projective
measurements.

\section{What the output distribution does not certify}
\label{sup:beyond}

\subsection{Correlators across the benchmark suite}
\label{sup:correlators-suite}

\begin{table}[htbp]
\caption{Two- and three-body correlators for every target, resolved by ladder.
The main text pools the two ladders; here they are separate. Values are the
correlators of the mean measured law over the $\genNreps$ FULL runs of each
cell, computed directly from the measured populations with no reconstruction of
any kind.}
\label{tab:sup-correlators-ab}
\centering\small
\begin{tabular}{llcccc}
\toprule
Target & & $\langle Z_0Z_1\rangle$ & $\langle Z_0Z_2\rangle$ & $\langle Z_1Z_2\rangle$ & $\langle Z_0Z_1Z_2\rangle$ \\
\midrule
D0 & target & $+0.000$ & $+0.000$ & $+0.000$ & $+0.000$ \\
   & ladder A & $+0.249$ & $-0.018$ & $+0.029$ & $-0.071$ \\
   & ladder B & $+0.185$ & $+0.022$ & $+0.091$ & $-0.063$ \\
\addlinespace
D1 & target & $-0.400$ & $-0.200$ & $+0.000$ & $+0.000$ \\
   & ladder A & $+0.128$ & $-0.156$ & $-0.084$ & $-0.125$ \\
   & ladder B & $+0.135$ & $-0.136$ & $-0.042$ & $-0.098$ \\
\addlinespace
D2 & target & $+0.280$ & $+0.100$ & $+0.060$ & $+0.040$ \\
   & ladder A & $+0.377$ & $+0.134$ & $-0.006$ & $-0.021$ \\
   & ladder B & $+0.340$ & $+0.099$ & $+0.046$ & $+0.048$ \\
\addlinespace
D3 & target & $-0.800$ & $-0.720$ & $+0.680$ & $+0.000$ \\
   & ladder A & $-0.327$ & $-0.375$ & $+0.161$ & $+0.084$ \\
   & ladder B & $-0.326$ & $-0.421$ & $+0.239$ & $+0.054$ \\
\addlinespace
D4 & target & $-0.720$ & $-0.440$ & $+0.320$ & $+0.000$ \\
   & ladder A & $-0.172$ & $-0.252$ & $+0.042$ & $-0.075$ \\
   & ladder B & $-0.148$ & $-0.325$ & $+0.028$ & $-0.008$ \\
\addlinespace
D5 & target & $+0.512$ & $-0.329$ & $+0.003$ & $+0.141$ \\
   & ladder A & $+0.320$ & $-0.285$ & $-0.070$ & $+0.160$ \\
   & ladder B & $+0.238$ & $-0.294$ & $-0.009$ & $+0.179$ \\
\addlinespace
D6 & target & $+0.126$ & $-0.353$ & $-0.400$ & $+0.348$ \\
   & ladder A & $+0.126$ & $-0.230$ & $-0.193$ & $+0.261$ \\
   & ladder B & $+0.169$ & $-0.376$ & $-0.275$ & $+0.346$ \\
\addlinespace
\bottomrule
\end{tabular}

\end{table}

Table~\ref{tab:sup-correlators-ab} is the ladder-resolved form of Table~\mref{tab:correlators_suite} of the
main text. It is included because a reader may reasonably want to check whether
the correlator discrepancies the main text reports are themselves
compilation-dependent, which the pooled table cannot show. They are, in the same
direction and of the same order as the fidelity contrasts, and we do not analyse
this further: the suite has one session per target, so a ladder-resolved
correlator analysis across targets inherits the alias of Section~\mref{subsec:suite-limits} intact. The
one correlator comparison in this paper that is free of that alias is the
intervention contrast of Section~\mref{subsec:int-correlator}, which is within a session and paired.

\subsection{The density matrices in the instrument record}
\label{sup:density}

\begin{table}[htbp]
\caption{Audit of the $8\times8$ \texttt{execution\_matrix} objects in the
SpinQuasar export, over all $\genDMn$ jobs of the October campaign, with the
seven duplicated job names resolved as described in
Section~\mref{subsec:density}. None of these quantities is used as a result in
either document;
Section~\mref{subsec:density-scope} of the main text says why.}
\label{tab:sup-density}
\centering\small
\begin{tabular}{ll}
\toprule
Property & Value \\
\midrule
Campaign jobs covered & 100 of 100 \\
Duplicated job names, resolved not dropped & 7 \\
Agreement of exported and deposited populations & $\le 7.5\times10^{-16}$ \\
Hermiticity, $\max|\rho-\rho^{\dagger}|$ & $5.7\times10^{-17}$ \\
Trace, $\max|\operatorname{tr}\rho-1|$ & $3.8\times10^{-15}$ \\
Smallest eigenvalue over all matrices & $-2.1\times10^{-16}$ \\
Diagonal against the reported populations & exact \\
Rank 3 / rank 4 / other & 59 / 41 / 0 \\
Rank of the paired simulation matrices & 1 throughout \\
Median purity $\operatorname{tr}\rho^{2}$ & $0.430$ \\
Closest pair of population vectors, $\TV$ & $0.048$ \\
\quad their largest off-diagonal difference & $0.097$ \\
Median $\langle\psi|\rho|\psi\rangle$ & $0.438$ over $[0.220,0.683]$ \\
Median classical fidelity, same runs & $0.897$ \\
Spearman $\rho$ between the two & $0.55$ \\
Ladder contrast, Control & $0.384$ vs $0.379$, $p=0.561$ \\
Ladder contrast, Reduced & $0.481$ vs $0.527$, $p=0.007$ \\
\bottomrule
\end{tabular}

\end{table}

Table~\ref{tab:sup-density} is the audit behind
Section~\mref{sec:density-record} of the main text.

The matrices are valid density matrices to numerical precision and their
diagonals are exactly the reported populations, so nothing in them contradicts
the measured counts. Their off-diagonal entries are not determined by those
counts: the two matrices whose population vectors are closest, at a total
variation of $\genDMcohTV$, differ by up to $\genDMcohDiff$ in the magnitude of
their off-diagonal entries. The object therefore carries information the counts
do not, which is why its provenance decides whether it can be used.

Every experimental matrix is rank-deficient, while the simulation matrices
supplied in the same records are rank $1$ throughout. The same audit over the
whole record --- all $\genDMrecN$ Grover--Rudolph jobs, \genDMrecSpan{} ---
gives ranks \genDMrecRanks{} (\genDMrecRankHist{}), never $1$ and never $8$,
with the vanishing eigenvalues at machine zero: the largest over the record is
$\genDMrecZeroMax$ (median $\genDMrecZeroMedian$) against a smallest surviving
eigenvalue of $\genDMrecNonzeroMin$, and only $\genDMrecMid$ of
$\genDMrecTotalEv$ eigenvalues fall between $10^{-12}$ and $10^{-3}$. Over the
same $\genDMrecDiagN$ jobs the diagonal reproduces the deposited populations to
the last bit.

We record that structure and draw nothing from it, and the provenance the
manufacturer has since specified accounts for it: a positivity correction is
applied before export, and such a correction produces exactly vanishing
eigenvalues as a matter of course \cite{ScholtenBlumeKohout2018}. The rank
deficiency is therefore a property of the processing and not of the underlying
physical ensemble.

The reason the implied state fidelities in this table are not reported as
results is the one Section~\mref{subsec:density-scope} gives. The
reconstruction draws its coefficients from the job's own readout across seven
settings, but consults simulated expectation values when selecting among
repeated coefficient estimates. Those simulated values come from the ideal
circuit, which is the same for the two compilations being compared, so the
estimator is not independent of the hypothesis a state-level contrast would be
testing.

\section{Experimental protocol and deposited artifacts}
\label{sup:protocol}

\subsection{Platform and software}

Both units are SpinQ Triangulum three-qubit liquid-state NMR processors
\cite{FengEtAl2022}, operated through the SpinQuasar client and the SpinQit
Python SDK. Circuits are expressed in the native gate set
$\{\Ry(\cdot),X,\mathrm{CNOT}\}$ and submitted job by job; the instrument
returns a population vector over the eight computational basis states, together
with the compiled gate list it executed and the $8\times8$ matrix of
Section~\ref{sup:density}.

\subsection{Campaign structure}

The benchmark campaign comprises $\genSessNtotal$ runs over seven targets,
$\genNreps$ in each of four cells per target: FULL with ladder~A, FULL with
ladder~B, $L0$ and $L01$, the truncated stages with ladder~A only. One target
was acquired per contiguous session in the order $\genSessOrder$, between
$\genSessFirst$ and $\genSessLast$ 2026, for $\genSessHours$ hours of
acquisition. Within each session the FULL-stage runs were acquired in
alternating blocks of five.

The intervention campaign comprises $100$ runs on D1 in a single session, four
cells of $\genNreps$ runs interleaved run by run and paired by repetition, with
the within-repetition order reversed on alternate repetitions
(Table~\ref{tab:sup-schedule}). Before any hardware job was submitted, each
circuit variant was simulated and checked against the analytic target, so that a
compilation mistake could not consume instrument time.

\subsection{Measurement conventions}

The bit-order convention was measured rather than assumed, by applying an $X$
gate to each qubit in turn with the others idle; the dominant outcomes were
\texttt{100}, \texttt{010} and \texttt{001}, confirming the canonical
most-significant-first assignment. All distributions in both documents are raw:
no readout mitigation is applied anywhere, for the experimental reason given in
Section~\mref{subsec:conventions} of the main text. Fidelities are reported as means of per-run values,
not as the fidelity of a pooled law; the two differ by up to $\genFidGapHi$ on
this campaign.

\subsection{Deposited artifacts}

\begin{itemize}[leftmargin=1.4em,itemsep=2pt]
\item Run-level records for the benchmark campaign: one row per hardware run
with the measured and reference distributions, the single-qubit marginals, the
stage-wise metrics and the circuit metadata.
\item Run-level records for the intervention campaign, in the same schema, with
the shot count, gate census and UTC timestamps recorded explicitly, plus the
acquisition script that produced them.
\item The readout-calibration session: run logs, the $8\times8$ confusion matrix,
and the raw, mitigated and partially-mitigated summaries behind Table~\mref{tab:readout} of the
main text.
\item Two SpinQuasar job database exports, the second superseding the first:
$1139$ jobs to $6$~October~2026, taken after the intervention campaign, and
$\genExportNjobs$ jobs spanning \genExportFrom{} to \genExportTo{}, taken
after the angle-representative campaign. Both carry the executed gate lists,
the measured populations and the density matrices. Its coverage of the benchmark campaign is
partial and Section~\mref{subsec:provenance} of the main text states exactly how.
\item The bit-order calibration script and its report.
\item The acquisition script for the angle-representative experiment of
Section~\mref{subsec:long-angle}, together with its $\genAngN$ runs per cell.
\item Two robustness analyses with their JSON output: the serial-dependence
treatment of the paired differences described in
Section~\ref{sup:intervention-unpaired}, and the Monte Carlo of the
independent per-gate angle-error model reported in
Section~\mref{subsec:disc-mechanism}.
\item The verification suite, which checks the exactness of the compiled
circuits, the A/B equivalence in the ideal model, the agreement between our
records and the instrument's, and the internal consistency of the pulse-level
tables. It needs neither hardware nor the vendor SDK.
\item The table-generating script for the CEU campaigns, which produces
their numerical tables and macros from the deposited run-level data.
The revised Shenzhen tables are author-supplied summaries; regeneration from
their underlying execution records remains pending.

\end{itemize}

For the Shenzhen campaign, the revised summaries and analysis conventions
are included here. Receipt of the underlying probability arrays and execution
scripts, independent regeneration of the summaries, and confirmation of their
public deposition remain outstanding, as stated in the data availability
statement.

\end{document}